\documentclass[12pt,a4paper,english,fleqn]{article}
\usepackage{mathptmx}
\usepackage{amssymb}
\usepackage{amsmath}
\usepackage[scaled=1.0]{helvet}
\usepackage[T1]{fontenc}
\usepackage[latin9]{inputenc}
\usepackage{fancyhdr}
\usepackage{graphicx}
\usepackage[authoryear]{natbib}

\numberwithin{equation}{section}

\usepackage[font=footnotesize,labelfont=bf]{caption}

\date{}

\usepackage{microtype}

\usepackage{subfig}
\AtBeginDocument{
  
}

\makeatother

\usepackage{babel}

\begin{document}

\title{Information equilibrium as an economic principle}

\author{Jason Smith%
\thanks{Associate Technical Fellow, The Boeing Company. P. O. Box 3707,
Seattle, Washington 98124. Email: jason.r.smith4@boeing.com.
}
}
\maketitle
\begin{abstract}
\noindent
A general information equilibrium model in the case of ideal information transfer is defined and then used to derive the relationship between supply (information destination) and demand (information source) with the price as the detector of information exchange between demand and supply. We recover the properties of the traditional economic supply-demand diagram. Information equilibrium is then applied to macroeconomic problems, recovering some common macroeconomic models in particular limits like the AD-AS model, IS-LM model (in a low inflation limit), the quantity theory of money (in a high inflation limit) and the Solow-Swan growth model. Information equilibrium results in empirically accurate models of inflation and interest rates, and can be used to motivate a ``statistical economics'', analogous to statistical mechanics for thermodynamics.

\noindent \medskip{}


\noindent \emph{Journal of Economic Literature Classification: }C00, E10, E30, E40\emph{.}
\end{abstract}
\noindent \newpage{}

\section{Introduction}\label{s:intro}

\noindent In the natural sciences, complex non-linear systems composed of large numbers of smaller subunits provide an opportunity to apply the tools of statistical mechanics and information theory. From this intuition Lee \cite{Smolin:2009} suggested a new discipline of statistical economics to study the collective behavior of economies composed of large numbers of economic agents.

A serious impasse to this approach is the lack of well-defined or even definable constraints enabling the use of Lagrange multipliers, partition functions and the machinery of statistical mechanics for systems away from equilibrium or for non-physical systems. The latter -- in particular economic systems -- lack e.g. fundamental conservation laws like the conservation of energy to form the basis of these constraints. In order to address this impasse, \cite{Fielitz:2014} introduced the concept of natural information equilibrium. They produced a framework based on information equilibrium and showed it was applicable to several physical systems. The present paper seeks to apply that framework to economic systems.

The idea of applying mathematical frameworks used in the physical sciences to economic systems is an old one; even the idea of applying principles from thermodynamics is an old one.  Willard Gibbs -- who coined the term "statistical mechanics" -- supervised Irving Fisher's thesis [\cite{Fisher:1892}] in which he applied a rigorous approach to economic equilibrium. Samuelson later codified the Lagrange multiplier approach to utility maximization commonly used in economics today.

The specific thrust of \cite{Fielitz:2014} is that it looks at how far you can go with the maximum entropy or information theoretic arguments without having to specify constraints. This refers to partition function constraints optimized with the use of Lagrange multipliers. In thermodynamics language it's a little more intuitive: basically the information transfer model allows you to look at thermodynamic systems without having defined a temperature (Lagrange multiplier) and without having the related constraint (that the system observables have some fixed value, i.e. equilibrium).

A word of caution before proceeding; the term "information" is somewhat overloaded across various technical fields. Our use of the word information differs from its more typical usage in economics, such as in ``information economics'' or ``perfect information'' in game theory. Instead of focusing on a board position in chess, we are assuming all possible board positions (even potentially some impossible ones such as those including three kings). The definition of information we use is the definition required when specifying a random chess board out of all possible chess positions, and it comes from Hartley and Shannon. It is a quantity measured in bits (or nats), and has a direct connection to probability. As stated in \cite{Shannon:1949}, ``information must not be confused with meaning''.

This is in contrast to Akerlof information asymmetry, for example, where knowledge (meaningful information) of the quality of a vehicle is better known to the seller than the buyer. We can see that this is a different use of the term information -- how many bits this quality score requires to store (and hence how many available `quality states' there are) is irrelevant to Akerlof's argument. The perfect information in a chess board $C$ represents $I(C) < 64 \log_{2} 13 \simeq 237$ bits; this quantity is irrelevant in an analysis of chess strategies in game theory (except as a practical limit to computation of all possible chess moves).

We propose the idea that information equilibrium should be used as a guiding principle in economics and organize this paper as follows. We will begin in Section \ref{s:infoeq} by introducing and deriving the primary equations of the information equilibrium framework, and proceed to show how the information equilibrium framework can be understood in terms of the general market forces of supply and demand. This framework will also provide a definition of the regime where market forces fail to reach equilibrium through information loss.

Since the framework itself is agnostic about the goods and services sold or the behaviors of the relevant economic agents, the generalization from widgets in a single market to an economy composed of a large number of markets is straightforward. We will describe macroeconomics in Section \ref{s:macro}, and demonstrate the effectiveness of the principle of information equilibrium both empirically an in derivations of standard macroeconomic models. In particular we will address the price level and the labor market where we show that information equilibrium leads to well-known stylized facts in economics. The quantity theory of money will be shown to be an approximation to information equilibrium when inflation is high, and Okun's law will be shown to follow from information equilibrium. Lastly, we establish in Section \ref{s:statecon} an economic partition function, define a concept of economic entropy and discuss how nominal rigidity and the so-called \emph{liquidity trap} in \cite{Krugman:1998} may be best understood as entropic forces for which there are no microfoundations.

\section{Information equilibrium}\label{s:infoeq}

\begin{figure}
\begin{centering}
\includegraphics[width=0.9\columnwidth]{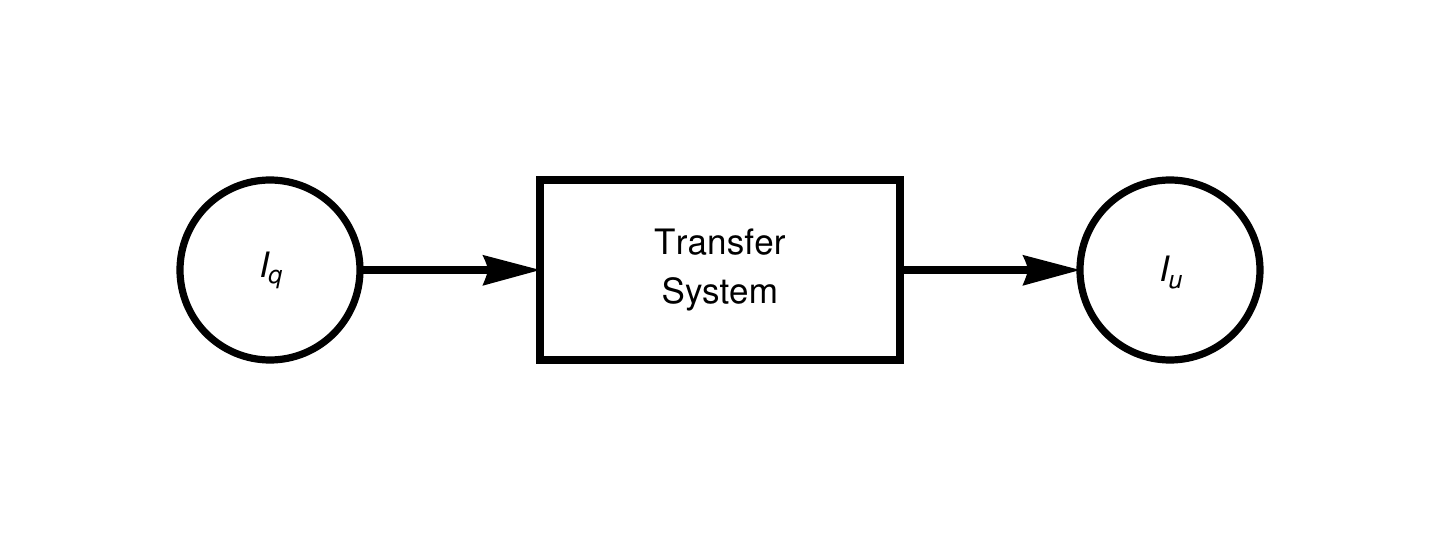}
\par\end{centering}
\caption{\label{fig:infotransfer}Information transfer from source to destination.}
\end{figure}

\noindent We will describe the economic laws of supply and demand as the result of an information transfer model. Much of the description of the information transfer model follows \cite{Fielitz:2014}. Following \cite{Shannon:1948} we have a system that transfers information\footnote{This follows the notation of one of the earlier versions of \cite{Fielitz:2014}. The German word for source is \textit{quelle}. We did not want to create confusion by using $S$ and $D$ for source and destination and then for supply and demand, since they appear in reverse order in the equations.} $I_q$ from a source $q$ to a destination $u$ (see Figure \ref{fig:infotransfer}). Any process can at best transfer complete information, so we know that $I_u \leq I_q$.

We will follow \cite{Fielitz:2014} and use the Hartley definition\footnote{The Hartley definition is equivalent to the Shannon definition for states with equal probabilities. As this definition enters into the information transfer index which is later taken as a free parameter, this distinction is not critical.} of information $I= K(s) n$ where $K(s) = K_{0} \log s$ where $s$ is the number of symbols and $K_{0}$ defines the unit of information (e.g. $1/\log 2$ for bits). If we take a measuring stick of length $|q|$ (process source) and subdivide it in to segments $\delta |q|$ (process source signal) then $n_q=|q|/\delta |q|$. In that case, the information transfer relationship $I_u \leq I_q$ becomes 
\begin{equation}
\frac{|u|}{\delta |u|} K_{u}(s_{u}) \leq \frac{|q|}{\delta |q|} K_{q}(s_{q})
\end{equation}
Let us define $k \equiv K_{q}(s_{q})/K_{u}(s_{u})$ which we will call the \emph{information transfer index} and rearrange so that
\begin{equation}
\frac{|u|}{\delta |u|} \leq k \frac{|q|}{\delta |q|}
\end{equation}
Compared to \cite{Fielitz:2014}, we have changed some of the notation, e.g. $|\Delta q|$ becomes $|q|$. We have set up the condition required by information theory for a signal $\delta |q|$ measured by the stick of length $|q|$ to be received as a signal $\delta |u|$ and measured by a stick of length $|u|$. These signals will contain the same amount of information if $I_u = I_q$.

Now we define a process signal detector that relates the process source signal $\delta |q|$ emitted from the process source $q$ to a process destination signal $\delta |u|$ that is detected at the process destination $u$ and delivers an output value:
\[
\space p \equiv \left(\frac{\delta |q|}{\delta |u|}\right)_{\mbox{\footnotesize detector}}
\]
If our source and destination are large compared to our signals ($n_q , n_u \gg 1$) we can take $\delta |q| \rightarrow d|q|$, we can re-arrange the information transfer condition:
\begin{equation}
\label{eq:itmequ}
p=\frac{d|q|}{d|u|} \leq k \frac{|q|}{|u|}
\end{equation}

In the following, we will use the notation\footnote{We can consider an information transfer model to be an `information preserving' morphism in category theory. The morphism itself is defined by the differential equation (\ref{eq:itmequ}), but we will label it with the detector $p$.} $p:q \rightarrow u$ to designate an information transfer model with source $q$, destination $u$ and detector $p$ for the general case where $I_u \leq I_q$, and use the notation $p:q \rightleftarrows u$ to designate an information equilibrium relationship where $I_{u} = I_{q}$. I will also occasionally use the notations $q \rightarrow u$ and $q \rightleftarrows u$ to designate an information transfer (information equilibrium) model without specifying the detector. Next, we derive supply and demand using this model.

\subsection{Supply and demand}\label{ss:snd}

\noindent At this point we will take our information transfer process and apply it to the generic economic problem of supply and demand. We will drop the absolute values and use positive quantities. In that case, we will identify the information transfer process source as the demand $D$, the information transfer process destination as the supply $S$, and the process signal detector as the price $P$. The price detector relates the demand signal $\delta D$ emitted from the demand $D$ to a supply signal $\delta S$ that is detected at the supply $S$ and delivers a price $P$. We translate Condition 1 in \cite{Fielitz:2014} for the applicability of our information theoretical description into the language of supply and demand:
\begin{quote}
\emph{Condition 1}: The considered economic process can be sufficiently described by only two independent process variables (supply and demand: $D, S$) and is able to transfer information.
\end{quote}
We are now going to solve the differential equation \ref{eq:itmequ}. But first we assume ideal information transfer $I_{S} = I_{D}$ such that:
\begin{equation}\label{eq:mainprice}
P= k \frac{D}{S}
\end{equation}
\begin{equation}\label{eq:main}
\frac{dD}{dS}= k \frac{D}{S}
\end{equation}
Note that Eq.~(\ref{eq:mainprice}) represents movement of the supply and demand curves where  $D$ is a ``floating-restriction'' information source in the language of \cite{Fielitz:2014}, as opposed to movement along the supply and demand curves where $D =D_0$ is a ``constant-restriction information source'', again in the language of \cite{Fielitz:2014}. The differential equation (\ref{eq:main}) can be solved by integration
\begin{eqnarray}
\int _{D_{ref}}^{D}\frac{d D'}{D'} & = & k \int_{S_{ref}}^{S} \frac{d S'}{S'} \\
 \log D -\log  D_{ref} & = & k \left( \log S -\log  S_{ref} \right) \\
\frac{D}{D_{ref}}  & = & \left( \frac{S}{S_{ref}} \right)^{k}
\end{eqnarray}
and we can then solve for the price using Eq.~(\ref{eq:mainprice})
\begin{eqnarray}
P & = & k \frac{D}{S}\\
 & = & k \frac{1}{S} D_{ref} \left( \frac{S}{S_{ref}} \right)^{k}\\
  & = & k \frac{1}{S} D_{ref} \frac{S}{S_{ref}} \left( \frac{S}{S_{ref}} \right)^{k-1}\\
  & = & k \frac{D_{ref}}{S_{ref}} \left( \frac{S}{S_{ref}} \right)^{k-1}
\end{eqnarray}
These equations represent the general equilibrium solution where $D$ and $S$ change in response to each other.

If we hold the information source or destination effectively constant, responding only slowly to changes in the other variable, we can describe `partial equilibrium' solutions that will lead to supply and demand diagrams. We will take $D =D_0$ to be a constant-restriction information source in the language of \cite{Fielitz:2014} and integrate the differential equation Eq.~(\ref{eq:main})
\[
\frac{1}{D_{0}}\int _{D_{ref}}^{D}dD' = k \int_{S_{ref}}^{S} \frac{1}{S} \, dS'
\]
We find
\begin{equation}\label{eq:demand1}
\Delta D=D-D_{ref}= k D_{0} \log \left(\frac{S}{S_{ref}}\right)
\end{equation}
Equation (\ref{eq:demand1}) represents movement along the demand curve, and the equilibrium price $P$ moves according to Eq.~(\ref{eq:mainprice}) based on the expected value of the supply and our constant demand source:
\begin{eqnarray}\label{eq:demandcurve1}
P & = &k \frac{D_{0}}{S}\\
\label{eq:demandcurve2}
\Delta D & = & k D_{0} \log \left(\frac{S}{S_{ref}}\right)
\end{eqnarray}
Equations (\ref{eq:demandcurve1},\ref{eq:demandcurve2}) define a demand curve. A family of demand curves can be generated by taking different values for $D_{0}$ assuming a constant information transfer index $k$.

Analogously, we can define a supply curve by using a constant information destination $S_{0}$ and follow the above procedure to find:
\begin{eqnarray}\label{eq:supplycurve1}
P & = & k \frac{D}{S_{0}}\\
\label{eq:supplycurve2}
\Delta S & = & \frac{S_{0}}{k} \log \left(\frac{D}{D_{ref}}\right)
\end{eqnarray}
So that equations (\ref{eq:supplycurve1}, \ref{eq:supplycurve2}) define a supply curve. Again, a family of supply curves can be generated by taking different values for $S_{0}$.

Note that equations (\ref{eq:demandcurve1},\ref{eq:demandcurve2}) and (\ref{eq:supplycurve1}, \ref{eq:supplycurve2}) linearize (Taylor series around $D=D_{ref}$ and $S=S_{ref}$)
\begin{eqnarray}
D & \simeq & D_{ref} + k D_{0} -S_{ref} P\label{eq:lineardemand}\\
S & \simeq & S_{ref}- \frac{S_{0}}{k} + \frac{S_{0}{}^2}{k^{2} D_{ref}}P\label{eq:linearsupply}
\end{eqnarray}
plus terms of order $P^2$ such that
\[
D \simeq \alpha -\beta P
\]
\[
S \simeq \gamma +\delta P
\]
where  $\alpha =D_{ref}+ k D_{0}$, $ \beta  = S_{ref}$ ,$ \gamma  = S_{ref}- S_{0}/k$ and $ \delta =S_{0}{}^2/(k^{2} D_{ref})$. This recovers a simple linear model of supply and demand (where you could add a time dependence to the price e.g. $ \frac{dP}{dt} \propto S - D $ to produce a simple dynamic model).

\begin{figure}[t]
\centering{}\subfloat[Supply and demand curves]{\centering{}\includegraphics[width=0.45\columnwidth]{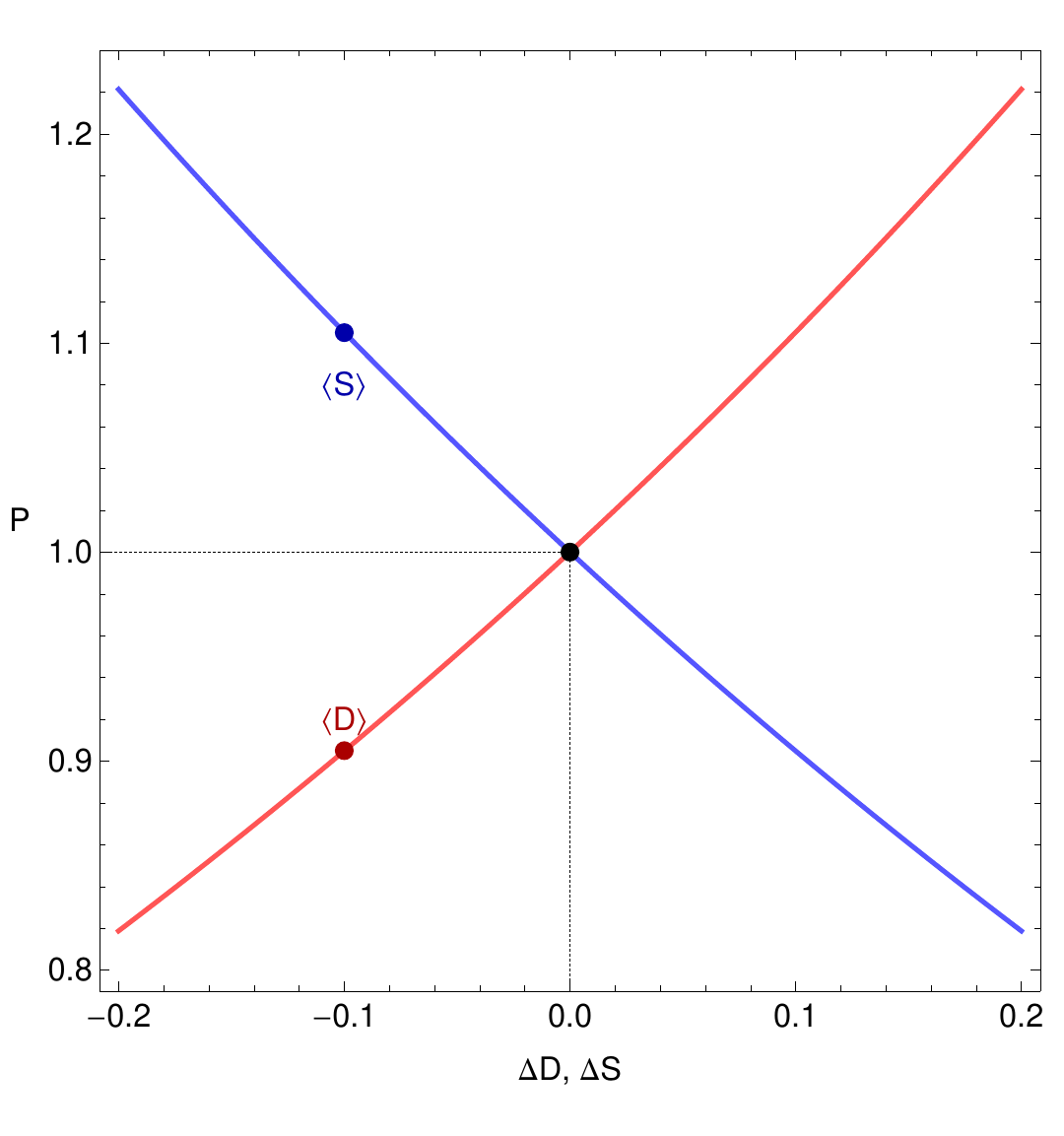}}\hfill{}\subfloat[Shift of the supply curve]{
\begin{centering}
\includegraphics[width=0.45\columnwidth]{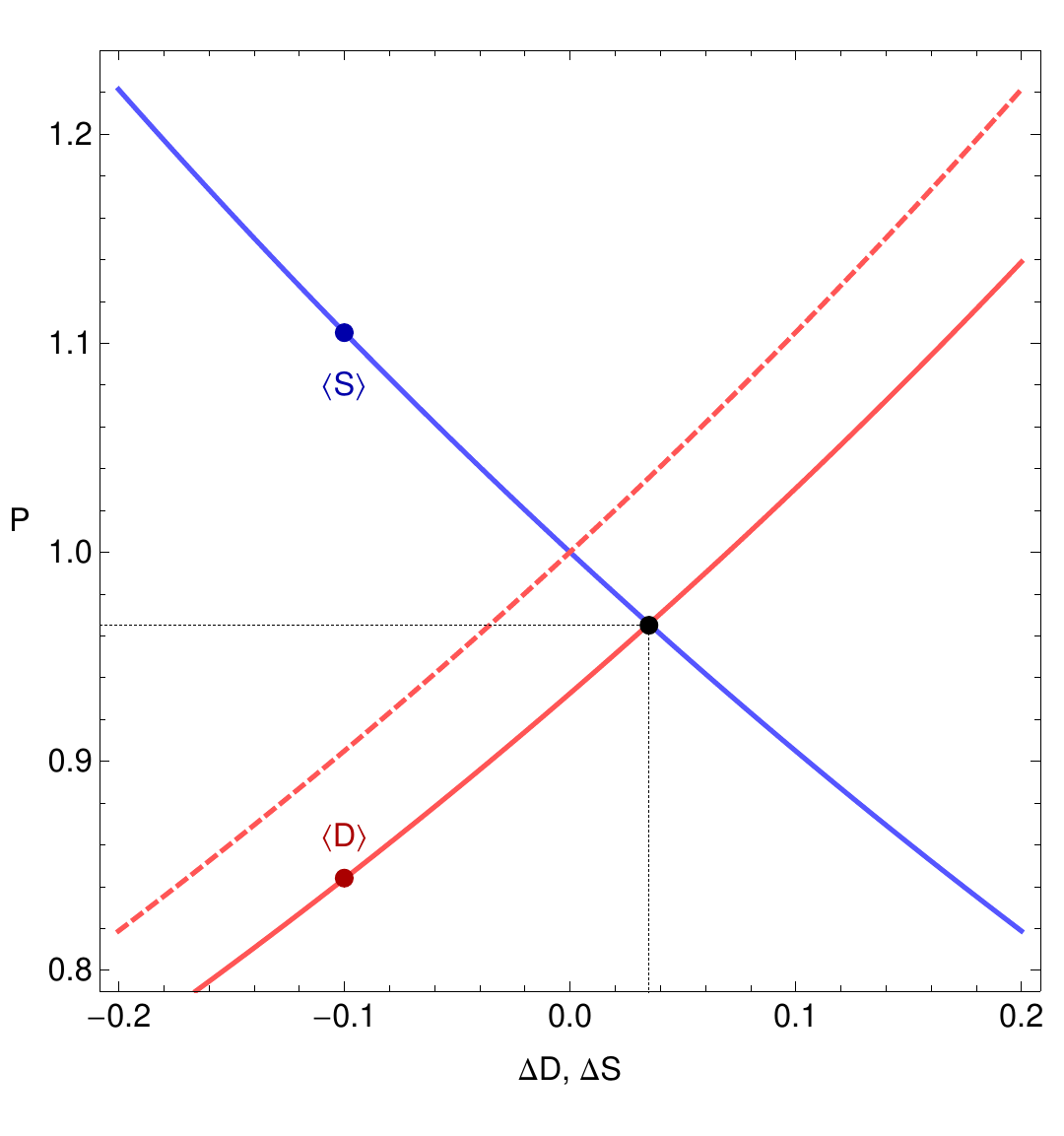}
\par\end{centering}
}\caption{\label{fig:sandd}Left: Supply and demand curves  Right: Shift of the supply curve resulting in a new lower equilibrium price.}
\end{figure}

We can explicitly show the supply and demand curves using equations (\ref{eq:demandcurve1},\ref{eq:demandcurve2}) and (\ref{eq:supplycurve1}, \ref{eq:supplycurve2}) and plotting price $P$ vs change in quantity $\Delta Q =\Delta S $ or $\Delta D$ in Figure \ref{fig:sandd}. In the figure we also show a shift in the supply curve (red) to the right. The new (lower) equilibrium price is the intersection of the new displaced supply curve and the unchanged demand curve.

If we use the linearized version of the supply and demand relationships Eqs.~(\ref{eq:lineardemand}, \ref{eq:linearsupply}) near the equilibrium price, we can find the (short run) price elasticities of demand and supply
\[
e^d = \frac{dD/D}{dP/P} =\frac{ D - D_0/k - D_{ref}}{D}
\]
Expanding around
\[
\Delta D=D-D_{ref}
\]
\[
e^d \simeq - \frac{k D_0}{D_{ref}} + O(\Delta D)
\]
And analogously
\[
e^s \simeq \frac{S_0}{k S_{ref}} + O(\Delta S)
\]
from which we could measure the information transfer index $k$.

There is a third way to solve Eq.~(\ref{eq:main}) where both supply and demand are considered to vary slowly (i.e.~be approximately constant). In that case the integral becomes
\[
\frac{1}{D_{0}}\int _{D_{ref}}^{D}dD' = \frac{k}{S_{0}} \int_{S_{ref}}^{S}  dS'
\]
If we define
\begin{eqnarray}
\Delta D & = & D-D_{ref}\nonumber\\
\Delta S & = & S-S_{ref}\nonumber
\end{eqnarray}
solving the integral shows us that the price is also constant
\begin{equation}
P = \frac{\Delta D}{\Delta S} = k \frac{D_{0}}{S_{0}}
\end{equation}

\subsection[Physical analogy]{Physical analogy with ideal gases}\label{ss:analogy}

\noindent In the original paper, \cite{Fielitz:2014} use the information transfer model to build the ideal gas law. This specific application gives us some analogies that are useful. In the model we have
\[
p = \frac{2}{f} \frac{E}{V}
\]
the pressure $p$ is the price $P$, volume $V$ is the supply $S$ and the energy\footnote{Substituting the energy in the formula you get $p V = N k_{B} T$} $E = (f/2) N k_{B} T$ is the demand $D$. The information transfer index contains the number of degrees of freedom $f$ in the ideal gas as well as the factor of $1/2$ that comes from the integral of a normal distribution in the derivation from statistical mechanics. In \cite{Fielitz:2014}, the general equilibrium solution corresponds to an isentropic process (and more generally, a polytropic process), while the partial equilibrium solution for the demand curve correspond to an isothermal process.

\subsection{Alternative motivation}\label{ss:alt}

\noindent We would like to provide an alternative and more macro- and micro-economic motivation of Eq.~(\ref{eq:main}) rooted in two economic principles: homogeneity of degree zero and marginalism. For example, according to Bennett \cite{McCallum:2004}, the quantity theory of money (QTM) is the macroeconomic observation that the economy obeys long run neutrality of money which is captured in the assumption of homogeneity constraints.  In particular, supply and demand functions will be homogeneous of degree zero, i.e. ratios of $D$ to $S$ such that if $D \rightarrow \alpha D$ and $S \rightarrow \alpha S$ then $g(D,S) \rightarrow \alpha^{0} g(D,S) = g(D,S)$. The simplest differential equation\footnote{We might consider this the most important term in an effective theory of supply and demand, analogous to effective field theory in physics where a full expansion would look like 
\[
\frac{dD}{dS} = c_{0} + c_{1} \frac{D}{S} + c_{2,0} \frac{D^{2}}{S^{2}} + \cdots + c_{2,1} \frac{D}{S} \frac{dD}{dS} + c_{2,2} D \frac{d^{2}D}{dS^{2}} + \cdots
\]} consistent with this observation is
\begin{equation}
\frac{dD}{dS} = k \frac{D}{S}
\end{equation}
\cite{Fisher:1892} looks at the exchange of some number of gallons of $A$ for some number of bushels of $B$ and states: "the last increment $dB$ is exchanged at the same rate for $dA$ as $A$ was exchanged for $B$". Fisher writes this as an equation on page 5:
\begin{equation}
\label{eq:fisherproto}
\frac{A}{B} = \frac{dA}{dB}
\end{equation}
Fisher notes that this marginalist argument was introduced by both Jevons and Marshall. Of course it is generally false. Many goods exhibit economies of scale, fixed costs or other effects so that either the last increments of $dA$ and $dB$ are cheaper (e.g. software) or more expensive (e.g. oil) than the first increments. The simplest way to account for this is by multiplying one side of Eq.~(\ref{eq:fisherproto}) by a constant. Thus we can say using information equilibrium as an economic principle enforces a generalized marginal thinking. The information equilibrium approach can also be interpreted as an application of information theory to Irving Fisher's measuring stick.

\section{Macroeconomics}\label{s:macro}

\noindent Since the information equilibrium framework depends on a large number of states for the information source and destination, it ostensibly would be better applied to the macroeconomic problem. Below we make a connection to some classic macroeconomic toy models and a macroeconomic relationship: AD-AS model, Okun's law, the IS-LM model, the Solow growth model, and the quantity theory of money. A summary of the models described in Section \ref{s:macro} appears in Appendix \ref{app:modeldetails}. The details of the \emph{Mathematica} codes used to fit the parameters are provides in Appendix \ref{app:codes}.

\subsection{AD-AS model}\label{ss:adasmodel}

\noindent The AD-AS model uses the price level $P$ as the detector, aggregate demand $N$ (NGDP) as the information source and aggregate supply $S$ as the destination, or $P:N \rightleftarrows S$, which immediately allows us to write down the aggregate demand and (short run) aggregate supply (SRAS) curves for the case of partial equilibrium.
\[
P = \frac{N_{0}}{k_{A} S_{ref}} \exp \left( - k_{A} \frac{\Delta N}{N_{0}} \right)
\]
\[
P = \frac{N_{ref}}{k_{A} S_{0}} \exp \left( + \frac{\Delta S}{k_{A} S_{0}} \right)
\]
Positive shifts in the aggregate demand curve raise the price level along with negative shifts in the supply curve. Traveling along the aggregate demand curve lowers the price level (more aggregate supply at constant demand).
\begin{figure}[t]
\centering{}\subfloat[AD-AS model]{\centering{}\includegraphics[width=0.48\columnwidth]{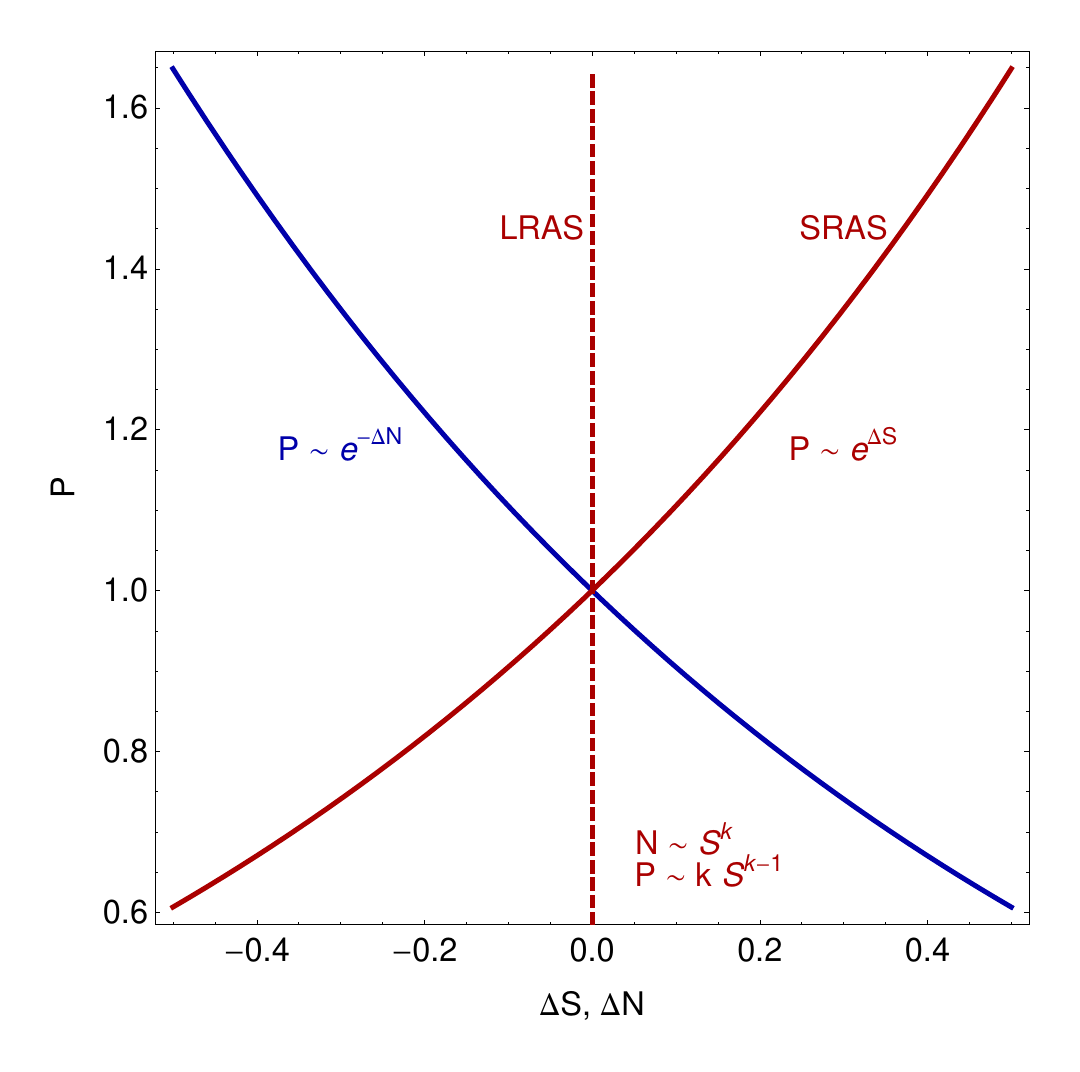}}\hfill{}\subfloat[Shift of the aggregate supply curve]{
\begin{centering}
\includegraphics[width=0.45\columnwidth]{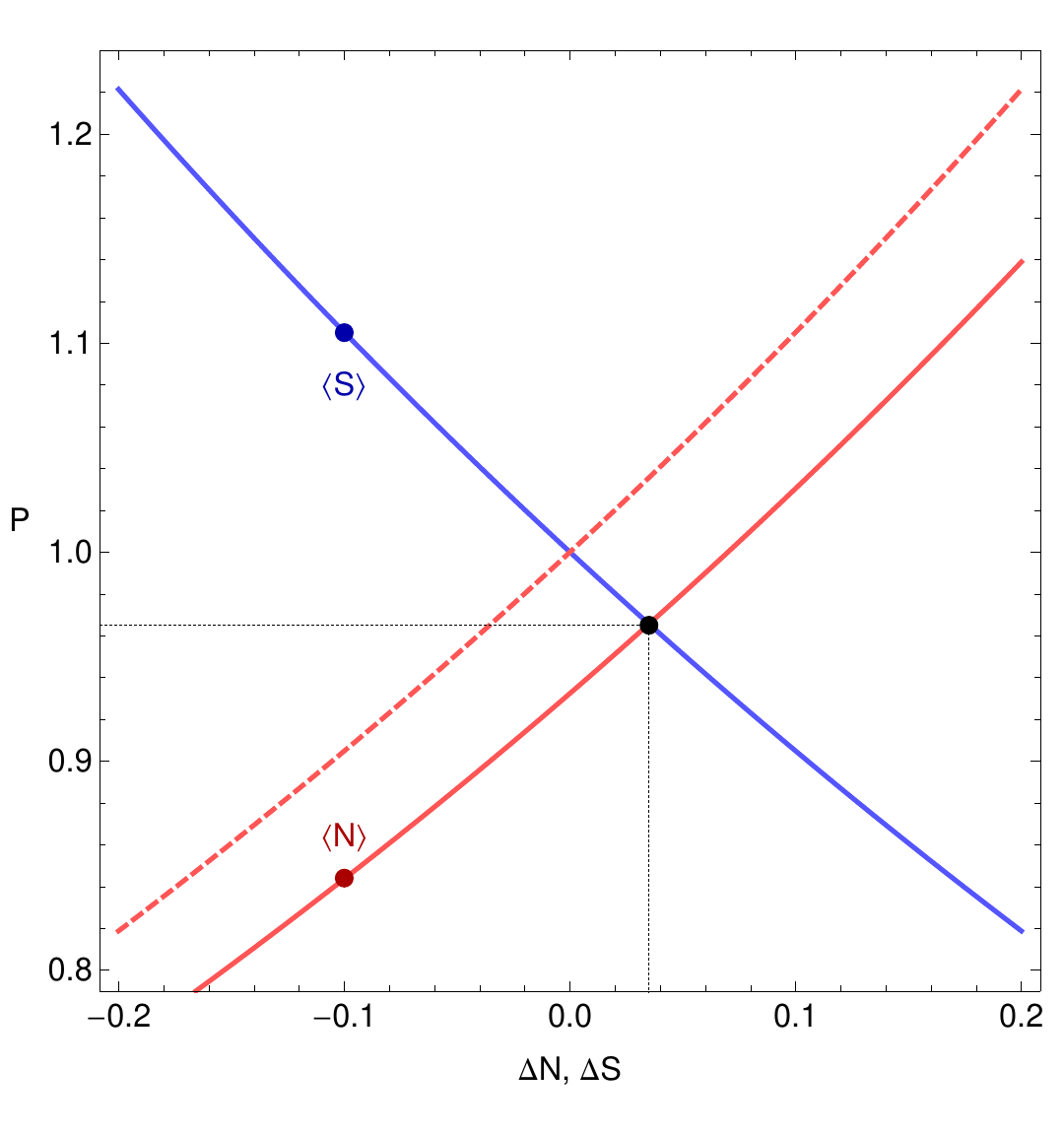}
\par\end{centering}
}\caption{\label{fig:adas}Left: AD-AS model with AD curve in blue, SRAS curve in red and LRAS curve as dashed red.  Right: Shift of the aggregate supply curve. The values $N = \langle N \rangle$ and $S = \langle S \rangle$ parameterize the supply and demand curves, respectively.}
\end{figure}
The long run aggregate supply (LRAS) curve would be vertical in Figure \ref{fig:adas} representing the general equilibrium solution
\[
\frac{N}{N_{ref}} = \left( \frac{S}{S_{ref}}\right)^{k_{A}}
\]
with price $P \sim S^{k_{A} - 1}$.

Another interesting result in this model is that it can be used to illuminate the role of money in macroeconomics as a tool of information mediation. If we start with the AD-AS model information equilibrium condition
\[
P \equiv \frac{dN}{dS} = k \; \frac{N}{S}
\]
we can in general make the following transformation using a new variable $M$ (i.e. money):
\begin{equation}\label{eq:moneymediation}
P = \frac{dN}{dM} \frac{dM}{dS} = k \; \frac{N}{M} \frac{M}{S}
\end{equation}
If we take $N$ to be in information equilibrium with the intermediate quantity $M$, which is in information equilibrium with $S$, i.e.
\[
P : N \rightleftarrows M \rightleftarrows S
\]
then we can use the information equilibrium condition
\[
\frac{dM}{dS} = k_{s} \; \frac{M}{S}
\]
to show that equation (\ref{eq:moneymediation}) can be re-written
\begin{eqnarray}
P & = & \frac{dN}{dM} \frac{dM}{dS} = k \; \frac{N}{M} \frac{M}{S}\\
& = &  \frac{dN}{dM}\left( k_{s} \; \frac{M}{S} \right) = k \; \frac{N}{M} \frac{M}{S}\\
& = & \frac{dN}{dM}  = \frac{k}{k_{s}} \; \frac{N}{M}\\
P & = & \frac{dN}{dM}  = k_{n} \; \frac{N}{M} \label{eq:adasQTM}
\end{eqnarray}
where we have defined $k_{n} \equiv k/k_{s}$. The solution to the differential equation (\ref{eq:adasQTM}) defines a quantity theory of money where the price level goes as
\[
\log P \sim ( k_{n} - 1 ) \log M
\]
We will discuss this more in Section \ref{ss:qtm} on the price level and inflation.

\subsection{Labor market and Okun's law}

\noindent The description of the labor market uses the price level $P$ as the detector, aggregate demand $N$ as the information source and total hours worked\footnote{You can also use the total employed $L$ as an information destination.} $H$ as the destination. We define the market $P:N \rightleftarrows H$ so that we can say:
\[
P = \frac{1}{k_{H}} \; \frac{N}{H}
\]
Re-arranging and taking the logarithmic derivative of both sides:
\begin{eqnarray}
H & = & \frac{1}{k_{H}} \; \frac{N}{P}\\
\frac{d}{dt} \log H & = & \frac{d}{dt} \log \frac{N}{P} - \frac{d}{dt} \log k_{H}\\
\frac{d}{dt} \log H & = & \frac{d}{dt} \log \frac{N}{P} - 0 = \frac{d}{dt} \log R
\end{eqnarray}
where $R$ is RGDP. The total hours worked $H$ (or total employed $L$) fluctuates with the change in RGDP growth. This is one form of Okun's law, from \cite{Okun:1962}. The model is shown in Figure \ref{fig:hoursworked}. The model parameters are listed in Appendix \ref{app:modeldetails}.
\begin{figure}[t]
\centering{}\includegraphics[width=0.8\columnwidth]{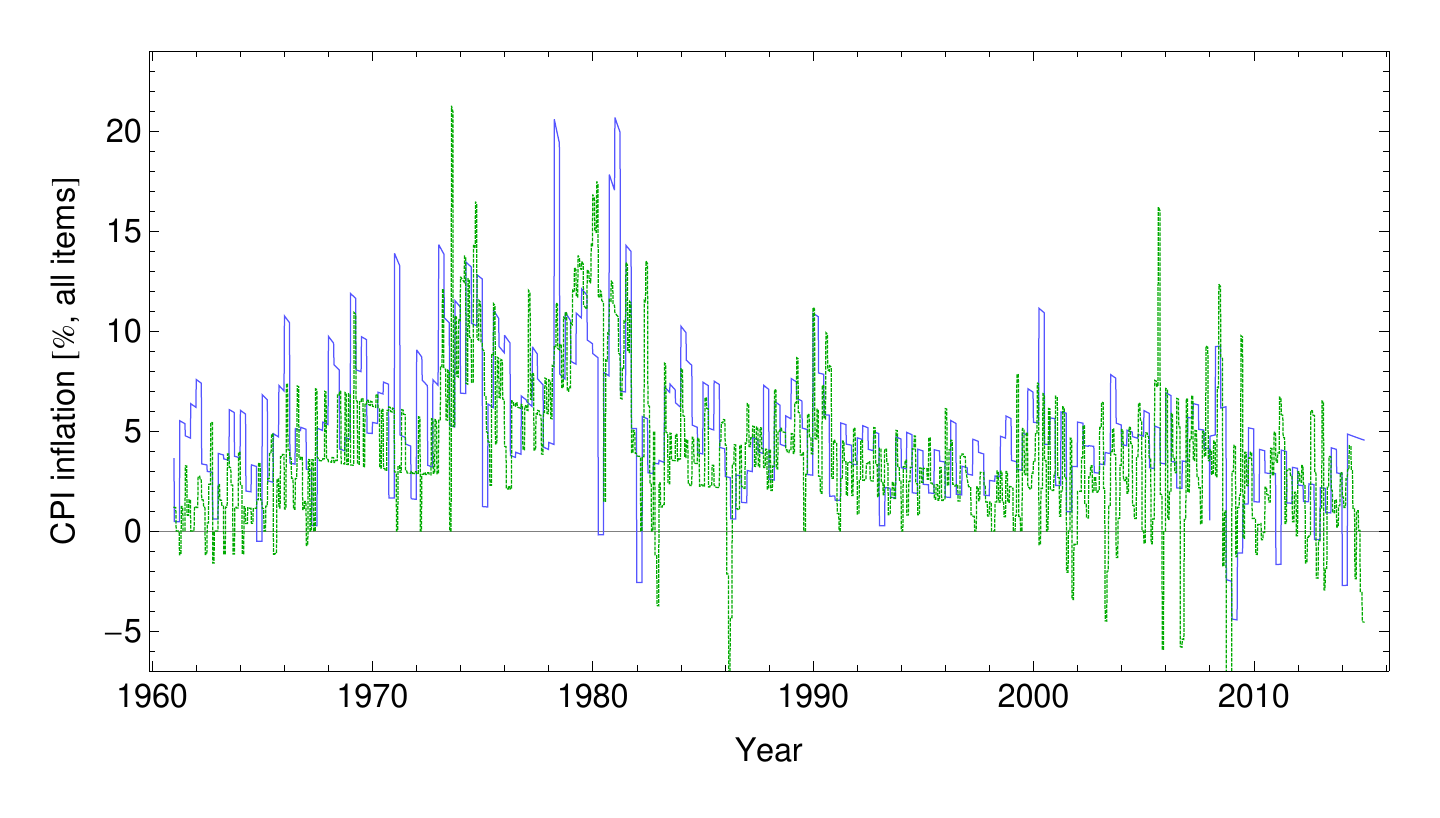}\caption{\label{fig:hoursworked}The model of US inflation using N = NGDP and total hours worked H is shown in blue. Inflation data (CPI all items) is in green.}
\end{figure}

\subsection{IS-LM and interest rates}\label{ss:islm}

\begin{figure}[t]
\centering{}\includegraphics[width=0.8\columnwidth]{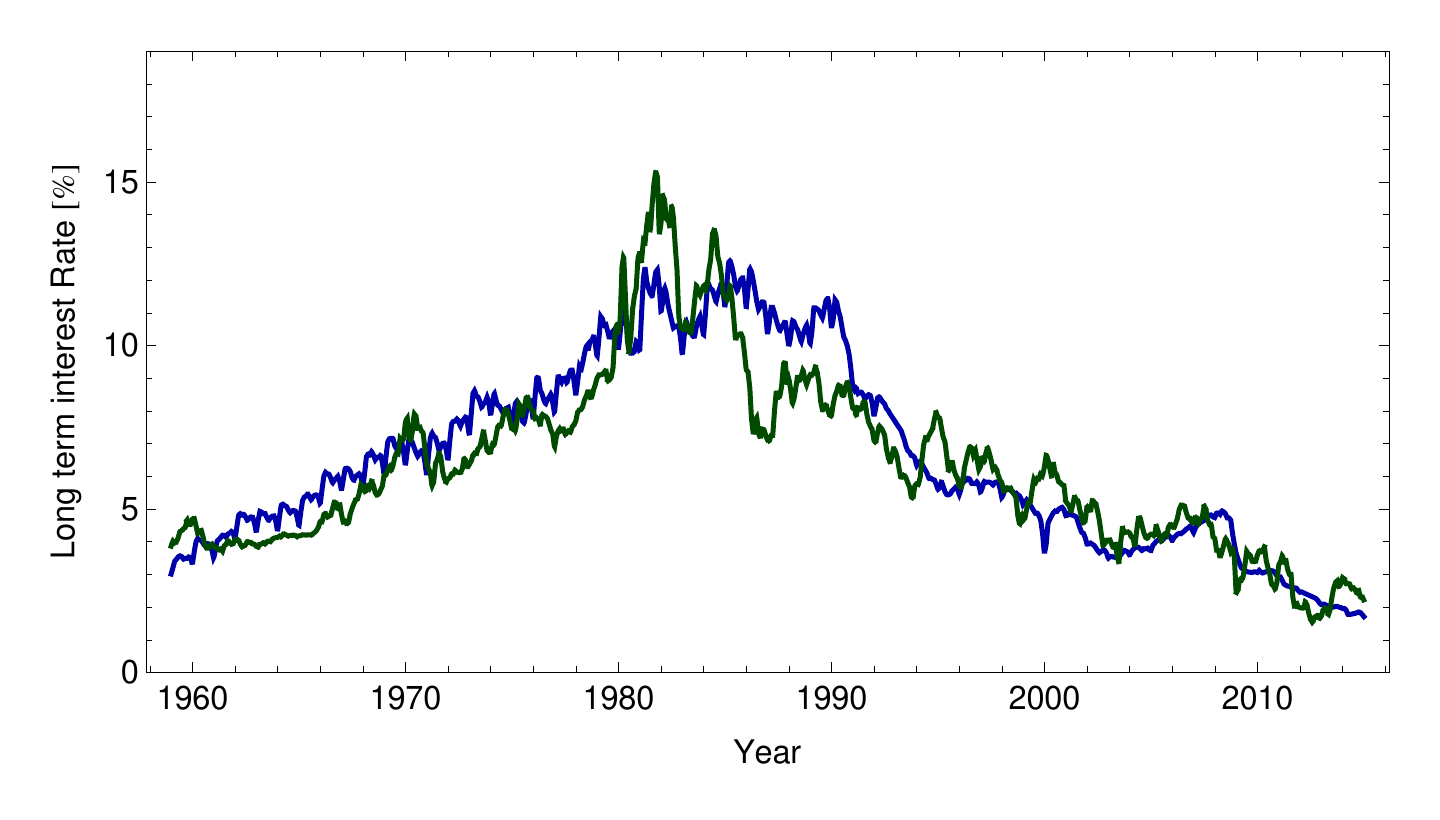}\caption{\label{fig:longrate}The model of US long term interest rate using N = NGDP and the monetary base minus reserves is shown in blue. The green dotted line is the long term interest rate data is from \cite{fred}; the data shown is the 10-year constant maturity rate, series GS10.}
\end{figure}

\noindent The classical Hicksian Investment-Savings Liquidity-Money Supply (IS-LM) model uses two markets along with an information equilibrium relationship. Let $p$ be the price of money in the money market (LM market) $p:N \rightleftarrows M$ where $N$ is aggregate demand and $M$ is the money supply. We have:
\[
p = \frac{1}{k_{p}} \; \frac{N}{M}
\]
We assume that the interest rate $i$ is in information equilibrium with the price of money $p$, so that we have the information equilibrium relationship $i \rightleftarrows p$ (no need to define a detector at this point). Therefore the differential equation is:
\[
\frac{di}{dp} = \frac{1}{k_{i}} \; \frac{i}{p}
\]
with solution (we will not need the additional constants $p_{ref}$ or $i_{ref}$):
\[
i^{k_{i}} = p
\]
And we can write:
\[
i^{k_i} = \frac{1}{k_{p}}  \; \frac{N}{M}
\]
Already this is fairly empirically accurate as we can see in Figure \ref{fig:longrate}.

We can now rewrite the money (LM) market and add the goods (IS) market as coupled markets with the same information source (aggregate demand) and same detector (interest rate, directly related to -- i.e. in information equilibrium with -- the price of money):
\begin{eqnarray}
i^{k_i} & : & N \rightleftarrows M \\
i^{k_i} & : & N \rightleftarrows S
\end{eqnarray}
\begin{figure}[t]
\centering{}\subfloat[IS market]{\centering{}\includegraphics[width=0.45\columnwidth]{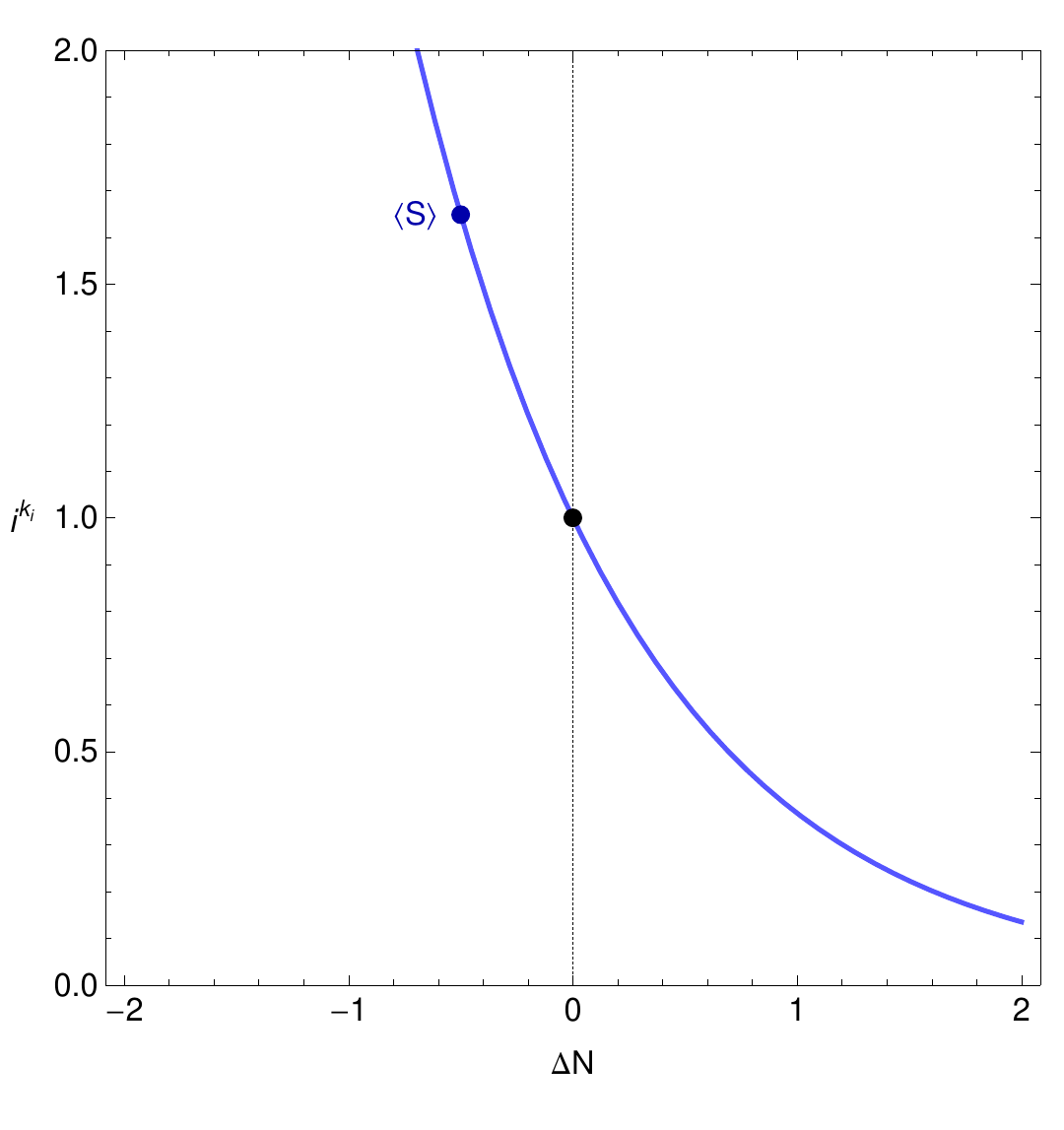}}\hfill{}\subfloat[LM market]{
\begin{centering}
\includegraphics[width=0.45\columnwidth]{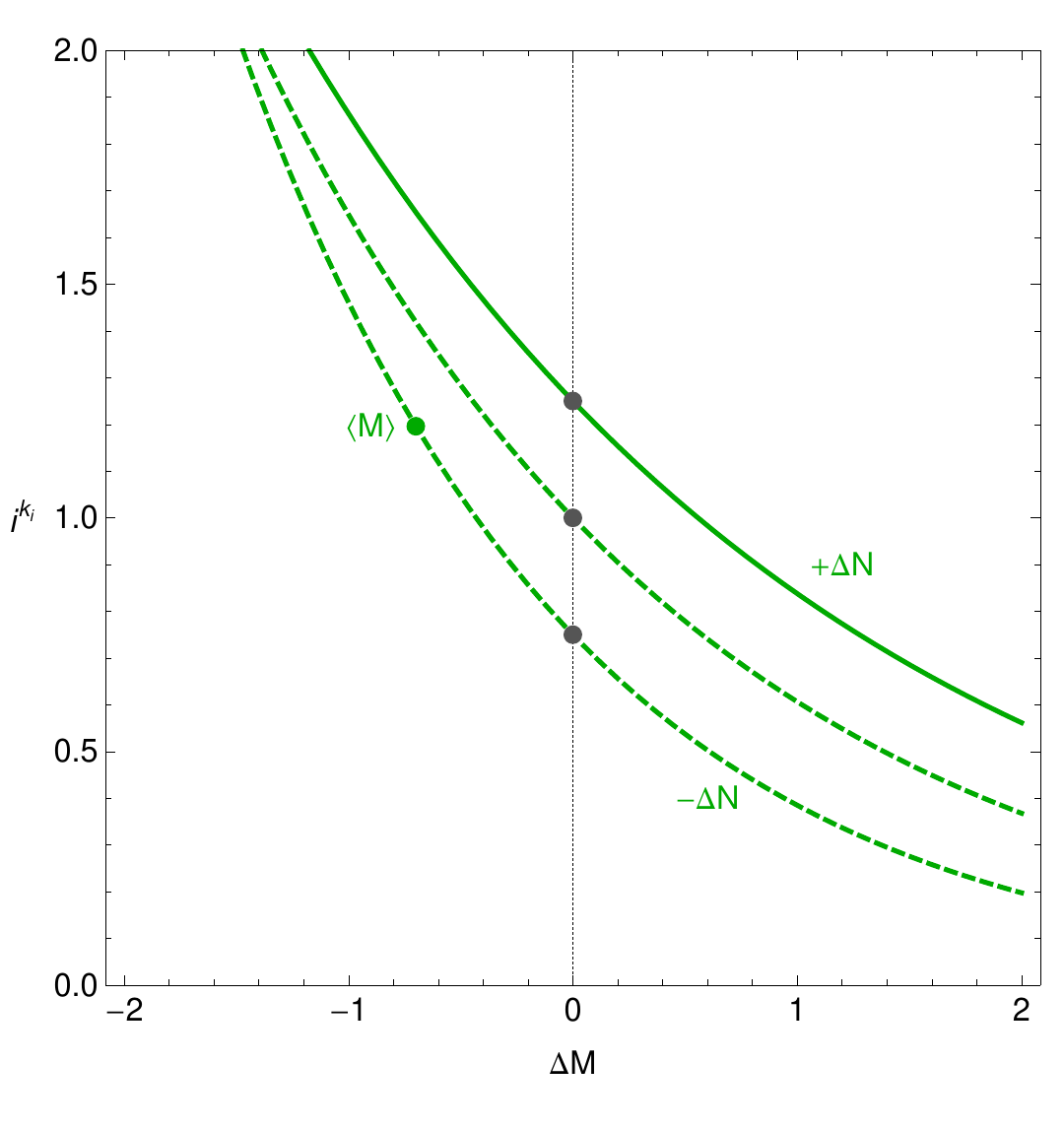}
\par\end{centering}
}\caption{\label{fig:isandlm}Left: IS market with IS curve in blue. Right: LM market with money demand curves in green. The three points represent the equilibrium solution as well as the equilibrium solutions after shifts of $\pm \Delta N$. The values $M = \langle M \rangle$ and $S = \langle S \rangle$ parameterize the money demand and IS curves, respectively.}
\end{figure}
where $S$ is the aggregate supply. Changes in the LM market manifest as increases in the money supply $M$ as well as shifts in the information source $N_{0} \rightarrow N_{0} + \Delta N$, so we write the LM curve as a demand curve Eqs.~(\ref{eq:demandcurve1}, \ref{eq:demandcurve1}) with shifts:
\[
i^{k_i} = \frac{N_{0} + \Delta N}{k_{p} M_{ref}} \exp \left( - k_{p} \frac{\Delta M}{N_{0} + \Delta N} \right)
\]
The IS curve can be straight-forwardly be written as the demand curve in the IS market:
\[
i^{k_i} = \frac{N_{0}}{k_{S} S_{ref}} \exp \left( - k_{S} \frac{\Delta N}{N_{0}} \right)
\]
This model assumes that $N$ does not move strongly with $M$, so only applies to a low inflation scenario. For high inflation, $N$ acquires a strong dependence on $M$ and the quantity theory of money in Section \ref{ss:qtm} becomes a more accurate description.
\begin{figure}[t]
\centering{}\includegraphics[width=0.8\columnwidth]{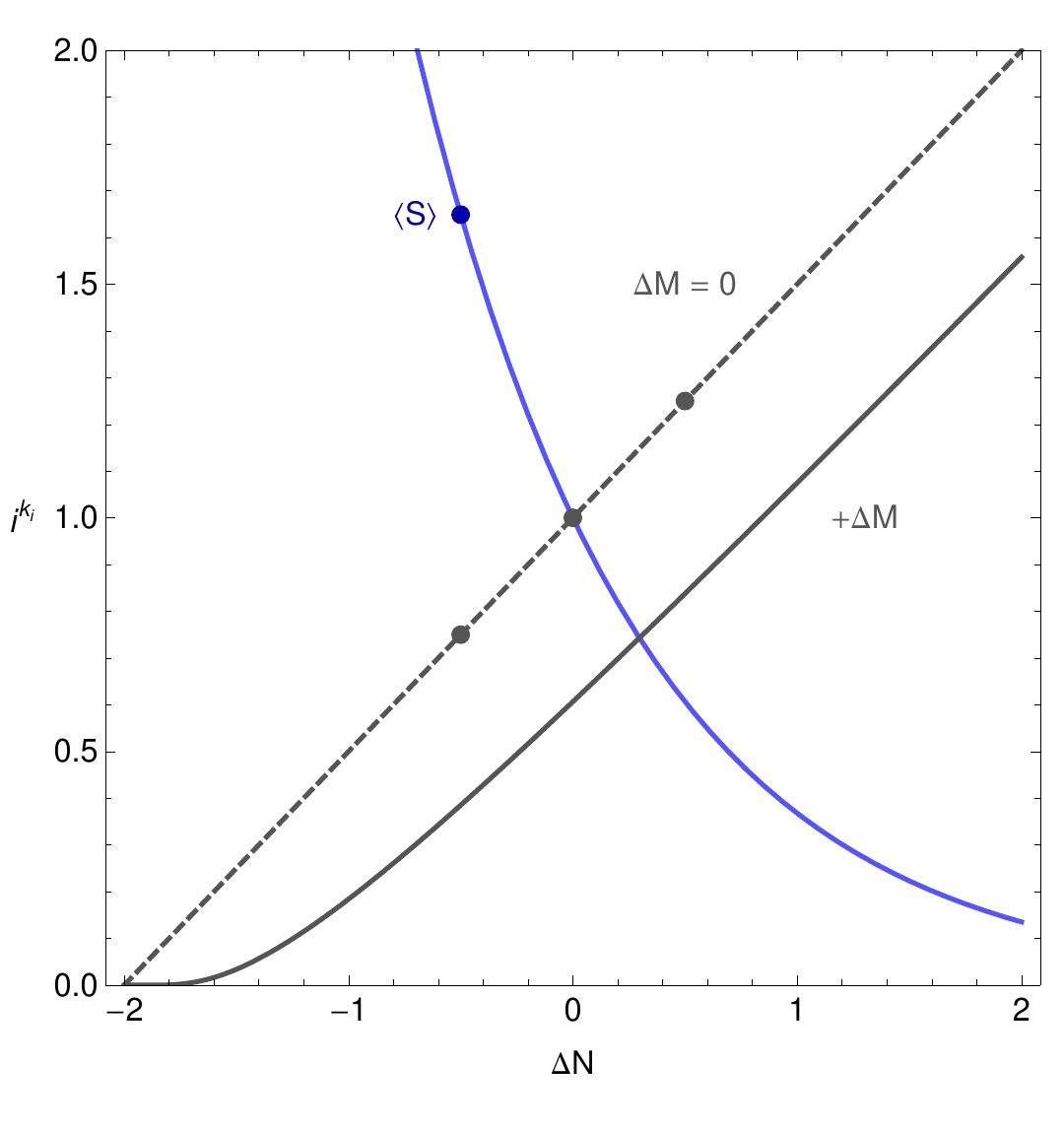}\caption{\label{fig:islm}The IS-LM model. The IS curve is in blue and the LM curve is gray. The three points on the $\Delta M = 0$ curve represent the same three points in Figure \ref{fig:isandlm}. An LM curve after a shift by $+\Delta M$.}
\end{figure}

\subsubsection{Long and short term interest rates}
\begin{figure}[t]
\centering{}\includegraphics[width=0.8\columnwidth]{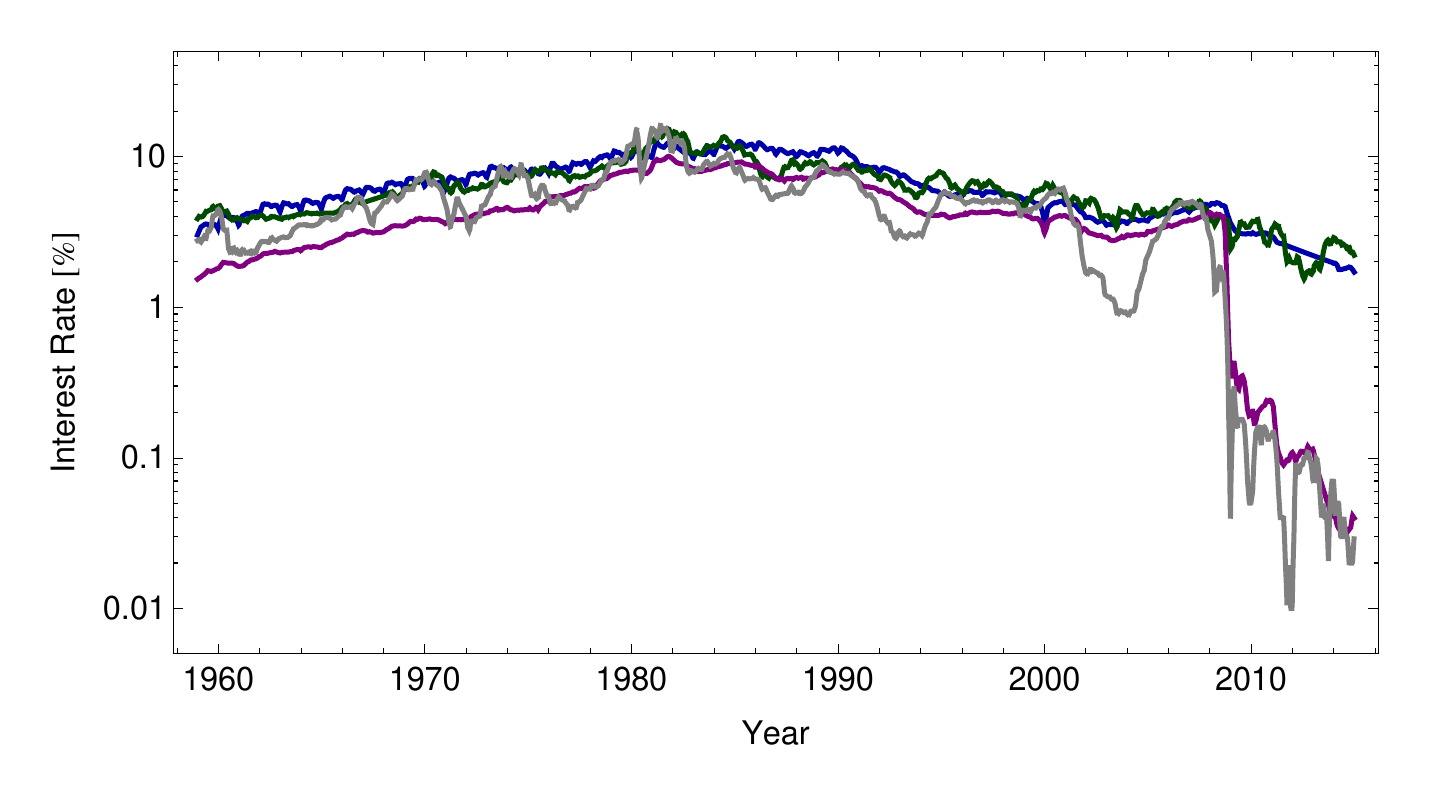}\caption{\label{fig:longshortrate}The model of US long and short interest rates. The short interest rate model using N = NGDP and M being the full monetary base(including reserves) is shown in purple. The long interest rate is as in Fig.~\ref{fig:longrate}. The gray dotted line is the short term interest rate data from \cite{fred}; the short interest rate data is taken to be the 3-month secondary market rate, series TB3MS.}
\end{figure}

\noindent The short term interest rate is empirically given by the same model with the same parameters (see Fig.~\ref{fig:longshortrate}); the difference is that the full monetary base including central bank reserves is used instead of just the currency component. These are \cite{fred} series AMBSL (call this variable $MB$) and MBCURRCIR, respectively. The full market for the long $i_{l}$ and short $i_{s}$ term interest rates would be:
\begin{eqnarray}
i_{l}^{k_i} & : & N \rightleftarrows M \\
i_{s}^{k_i} & : & N \rightleftarrows MB
\end{eqnarray}
where $k_{i_{l}} = k_{i_{s}} = k_{i}$ and $k_{p_{l}} = k_{p_{s}} = k_{p}$ (\textit{i.e.} the parameters for both models are the same).

The theoretical reason both the long and short term interest rate are given by the same model simply by exchanging currency (monetary base minus reserves) for the full monetary base (including reserves) is not immediately obvious. As the relationship was observed in empirical data, we can only provide a hand-waving argument based on the properties of central bank reserves (which are purely electronic) as opposed to currency which manifests as physical pieces of paper. Reserves may be seen as temporary by the market - they only exist in the short run. Therefore they need to be included as part of the supply of so-called high powered money for short term interest rates. Physical currency in circulation may be seen as more permanent by the market, and therefore represent the proper supply of high powered money for long term interest rates. This argument is speculative and involves the expected path of the monetary base, something not empirically measurable.

\subsubsection{Assumptions in the IS-LM model}

\noindent One useful property of the information equilibrium approach is that is makes explicit several assumptions in the IS-LM model.
\begin{itemize}
\item It is a partial equilibrium model and we use the partial equilibrium solutions to the information equilibrium equation Eq.~(\ref{eq:itmequ}).
\item No distinction is made between real and nominal quantities (all quantities are treated as nominal). Since we have partial equilibrium, $N$ is assumed to be slowly varying which implies that if $N = PY$, $P$ must be slowly varying unless $P$ and $Y$ conspire to make $N$ slowly varying.
\item If the price of money is scaled by a constant factor $p \rightarrow \alpha p$, the only change to the model is a change in the value of the constant $k_{p} \rightarrow k_{p}/\alpha$.
\end{itemize}

\subsection{Price level and inflation}\label{ss:qtm}

\noindent Let us begin our discussion of the price level with the market $P : N \rightleftarrows M$ described in the AD-AS model in section \ref{ss:adasmodel} with $N$ being NGDP, the information source, and $M$ being the monetary base minus reserves and return to the differential equation (\ref{eq:itmequ}). Assuming ideal information transfer we have
\begin{equation}\label{eq:diffnm}
P = \frac{dN}{dM} = k \; \frac{N}{M}
\end{equation}
Let us allow $k = k(N,M)$ to be a slowly varying function of $N$ and $M$, i.e.
\begin{equation}\label{eq:slowk}
\frac{\partial k}{\partial N} \mbox{ , } \frac{\partial k}{\partial M} \approx 0 
\end{equation}
We can approximately solve the differential equation (\ref{eq:diffnm}) by integration such that
\begin{eqnarray}
\int_{N_{0}}^{N} \frac{dN'}{N'} & \approx & k(N, M) \int_{M_{0}}^{M} \frac{dM'}{M'}\\
\rightarrow \; \frac{N}{N_{0}} & = & \left(\frac{M}{M_{0}} \right)^{k(N,M)}
\end{eqnarray}
so that, using Eq.~(\ref{eq:diffnm}) again, we obtain the price level as a function of $N$ and $M$
\begin{equation}
P(N,M) \approx \alpha \; k(N,M) \left(\frac{M}{M_{0}} \right)^{k(N,M) - 1}
\end{equation}
where $\alpha$ is an arbitrary constant (because the normalization of the price level is arbitrary).

Now the information transfer index $k$ is related to the number of symbols $s_N$, $s_{M}$ used by the information source and information destination, specifically:
\begin{equation}
k = \frac{K_{0} \log s_{N}}{K_{0} \log s_{M}} = \frac{\log s_{N}}{\log s_{M}}
\end{equation}
Let us posit a simple model where $s_{N}$ and $s_{M}$ are proportional to $N$ and $M$
\begin{eqnarray}\label{eq:defkappa}
s_{N} & = & N/(\gamma M_{0})\\
s_{M} & = & M/(\gamma M_{0})\\
k(N, M) & = & \frac{\log N/(\gamma M_{0}) }{\log M/(\gamma M_{0}) }
\end{eqnarray}
where we have introduced the new\footnote{We have simply traded the parameter degree of freedom $k$ in the constant information transfer index version of the model for $\gamma$; we have not increased the number of parameters in the model.} dimensionless parameter $\gamma$. This functional form meets the requirement that $k(N,M)$ is slowly varying with $N$ and $M$:
\begin{eqnarray}\label{eq:slowkappa}
\frac{\partial k}{\partial N} & = & \frac{\gamma M_{0}}{N \log M/(\gamma M_{0})} \approx 0 \\
\frac{\partial k}{\partial M} & = & - \frac{\log N/(\gamma M_{0})}{M (\log M/(\gamma M_{0}))^{2}} \approx 0
\end{eqnarray}
for $N,M \gg 1$. The rationale for introducing such a model for a changing information transfer index $k$ is that the units of $N$ and $M$ are the same: the national unit of account. Therefore the information content of \textit{e.g.} \$ 1 billion of nominal output depends on the size of the monetary base -- and \textit{vice versa}, and so we should expect $k = k(N, M)$. However, we will see in Section \ref{s:statecon} that this functional form is a good approximation to the case where we consider $n \gg 1$ markets with a distribution of constant values of $k$, meaning $k = k(N, M)$ effectively describes \emph{emergent} properties of the macroeconomy. There is an additional benefit of introducing this functional form and constant $\gamma$ that may assist in cross-national comparisons that we discuss in Appendix \ref{app:sigmakappa}.

\begin{figure}[t]
\centering{}\includegraphics[width=0.8\columnwidth]{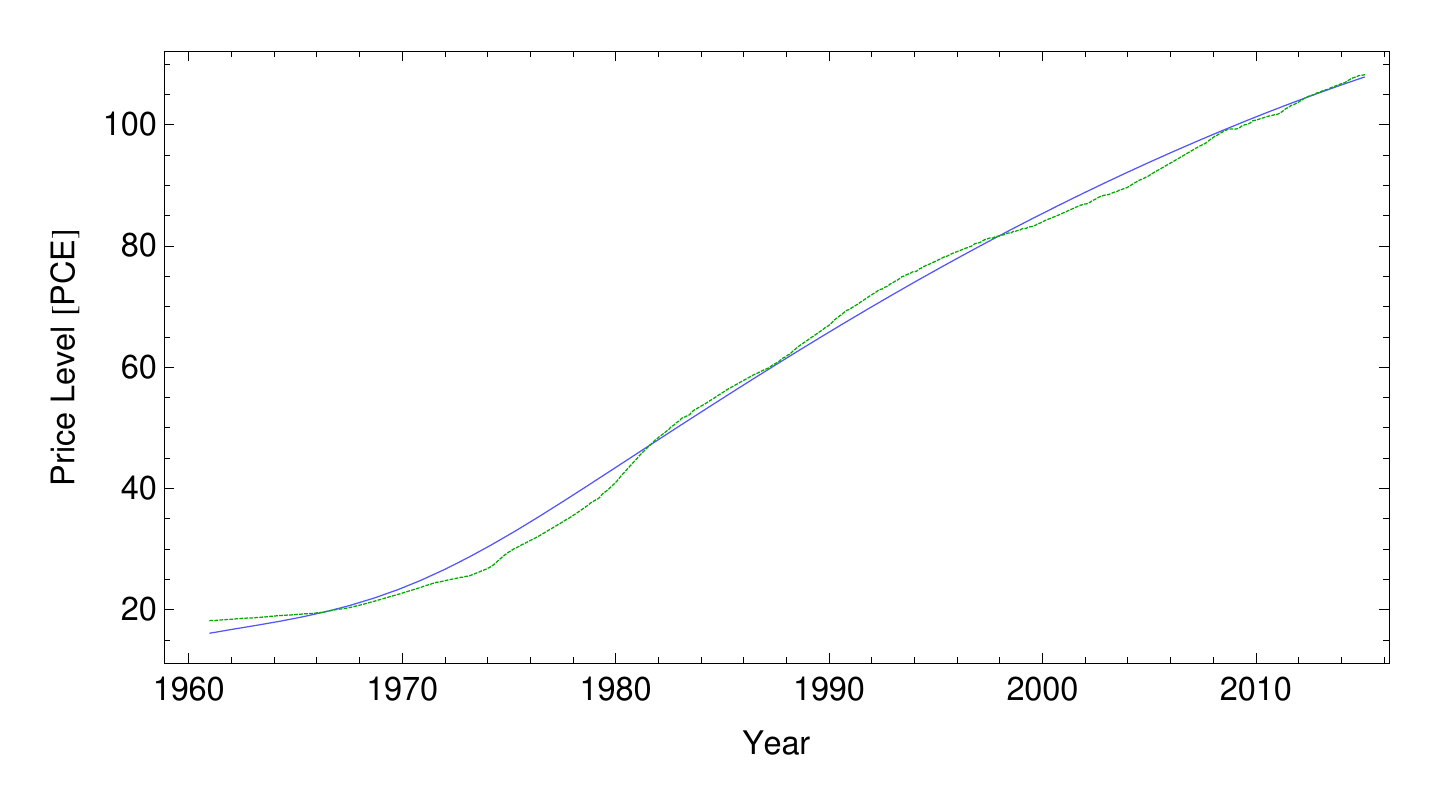}\caption{\label{fig:pcelevel}The model of US price level with N being NGDP and the M being monetary base minus reserves (MBCURRCIR) is shown in blue. Price level data (core PCE, with 2011 = 100) is in green.}
\end{figure}

\begin{figure}[t]
\centering{}\includegraphics[width=0.8\columnwidth]{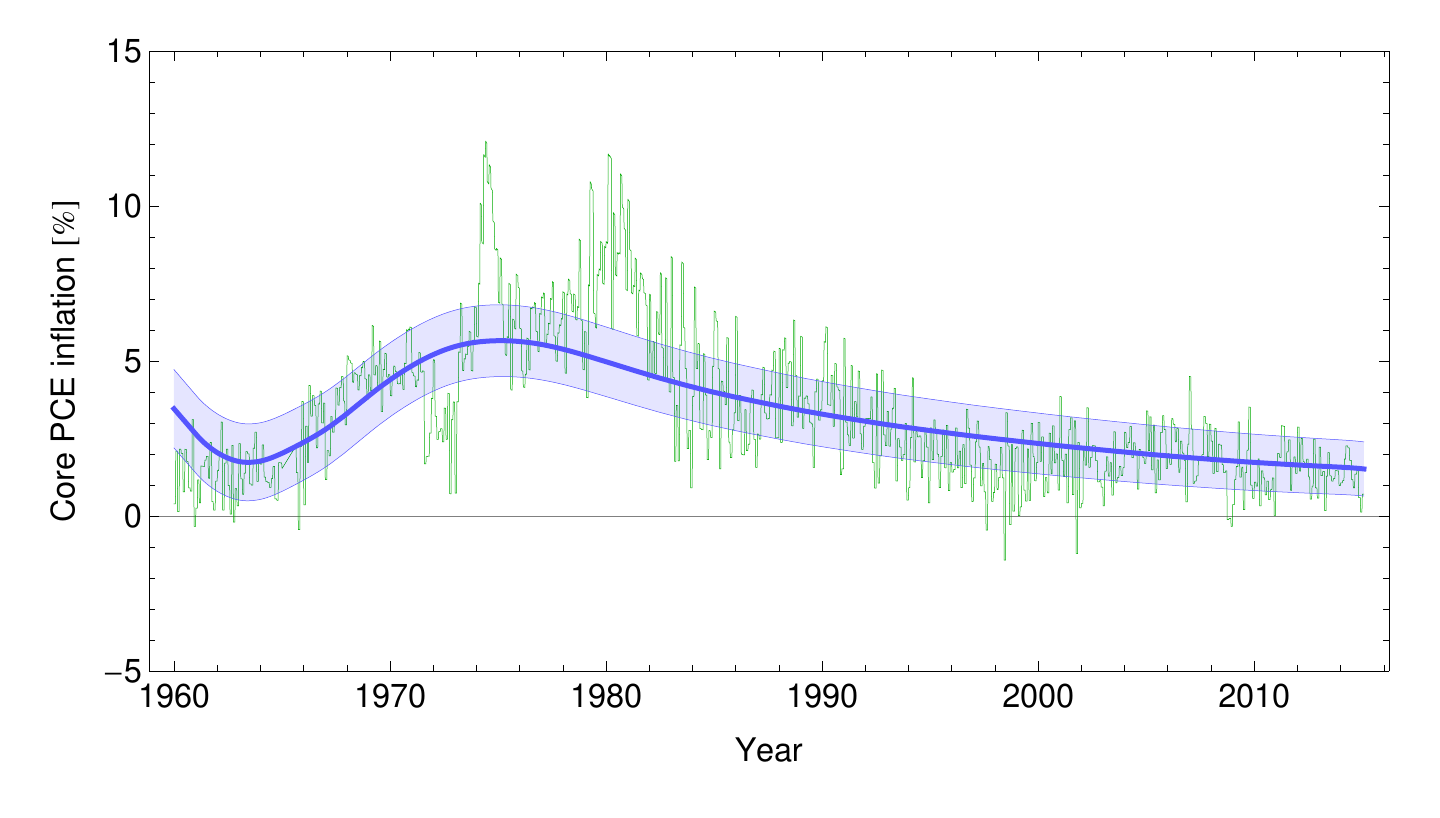}\caption{\label{fig:pceinf}The model of US inflation using N = NGDP and the monetary base minus reserves is shown in blue. Inflation data (core PCE) is in green. The blue bands represent 1-$\sigma$ error bands for the residuals.}
\end{figure}

The full price level model is
\begin{equation}\label{eq:pricelevel}
P(N,M) \approx \alpha \; \frac{\log N/(\gamma M_{0}) }{\log M/(\gamma M_{0}) } \left(\frac{M}{M_{0}} \right)^{\frac{\log N/(\gamma M_{0}) }{\log M/(\gamma M_{0}) } - 1}
\end{equation}
with free dimensionless parameters $\alpha$ and $\gamma$ along with $M_{0}$, which has dimensions of currency. If we fit these parameters using data from \cite{fred} for $P$ being so-called core price level of Personal Consumption Expenditures (PCE price level, less food and energy, series PCEPILFE), $N$ being nominal gross domestic product (series GDP), and $M$ being the currency component of the monetary base (series MBCURRCIR), performing a LOESS smoothing (of order 2, with smoothing parameter 1.0, see Appendix \ref{app:codes}) on the inputs $N$ and $M$ we arrive at Figures \ref{fig:pcelevel} and \ref{fig:pceinf}. The empirical accuracy of the model is on the order of the $P^{*}$ model of \cite{pstar:1989} (see Appendix \ref{app:modeldetails} for fit parameters).

If we look at Eq.~(\ref{eq:pricelevel}) we can see that when $k = 2$, we have
\begin{eqnarray}
P(N,M) & = & 2 \alpha \;  \left(\frac{M}{M_{0}} \right)^{2 - 1} \\
P & \sim & M 
\end{eqnarray}
so that price level grows proportionally with the monetary base, the essence of the quantity theory of money. Additionally, when $k = 2$ we have, using Eq.~(\ref{eq:defkappa}),
\begin{eqnarray}
2 (\log M - \log \gamma M_{0}) & = & \log N - \log \gamma M_{0}\\
M^{2} & \sim & N
\end{eqnarray}
If we use the fact that $M, N \gg 1$. If we take $M$ and $N$ to be exponentially growing with growth rates $m$ and $n$ (i.e. $M \sim \exp m t $), respectively, $2 m = n$. In general, we have (introducing the inflation rate $\pi$)
\begin{eqnarray}
\pi  & \simeq & (k - 1) m\\
n & \simeq & k m
\end{eqnarray}
Defining a real growth rate $r \equiv n - \pi$, then for large $k \gg 1$ we have
\begin{equation}
\frac{n}{\pi} = \frac{k m}{(k - 1) m} \simeq 1 \simeq \frac{\pi + r}{\pi} \sim \frac{\pi}{\pi}
\end{equation}
which implies large $k$ means high inflation. In contrast, $k \approx 1$ means that $\pi \approx 0$. When $k \approx 1$, the IS-LM model becomes a better approximation since changes in $M$ do not result in strong changes in the price level $P$ since $P \sim M^{k - 1} \sim M^{0} = 1$. We will discuss this more in Section \ref{ss:entropic}.

\subsection{Solow-Swan growth model}

\noindent Let us assume two markets $p_{1}:N \rightleftarrows K$ and $p_{2}:N \rightleftarrows L$:
\begin{eqnarray}\label{eq:solow}
\frac{\partial N}{\partial K} & = & k_{1} \; \frac{N}{K} \\
\frac{\partial N}{\partial L} & = & k_{2} \frac{N}{L}
\end{eqnarray}
The economics rationale for equations (\ref{eq:solow}) are that the left hand sides are the marginal productivity of capital/labor which are assumed to be proportional to the right hand sides -- the productivity per unit capital/labor. In the information transfer model, the relationship follows from a model of aggregate demand sending information to aggregate supply (capital and labor) where the information transfer is ``ideal'' i.e. no information loss. The solutions are:
\[
N(K, L) \sim f(L) K^{k_{1}}
\]
\[
N(K, L) \sim g(K) L^{k_{2}}
\]
and therefore we have
\begin{equation}\label{eq:solow2}
N(K, L) = A  K^{k_{1}} L^{k_{2}}
\end{equation}

\begin{figure}[t]
\centering{}\subfloat[Output level]{\centering{}\includegraphics[width=0.45\columnwidth]{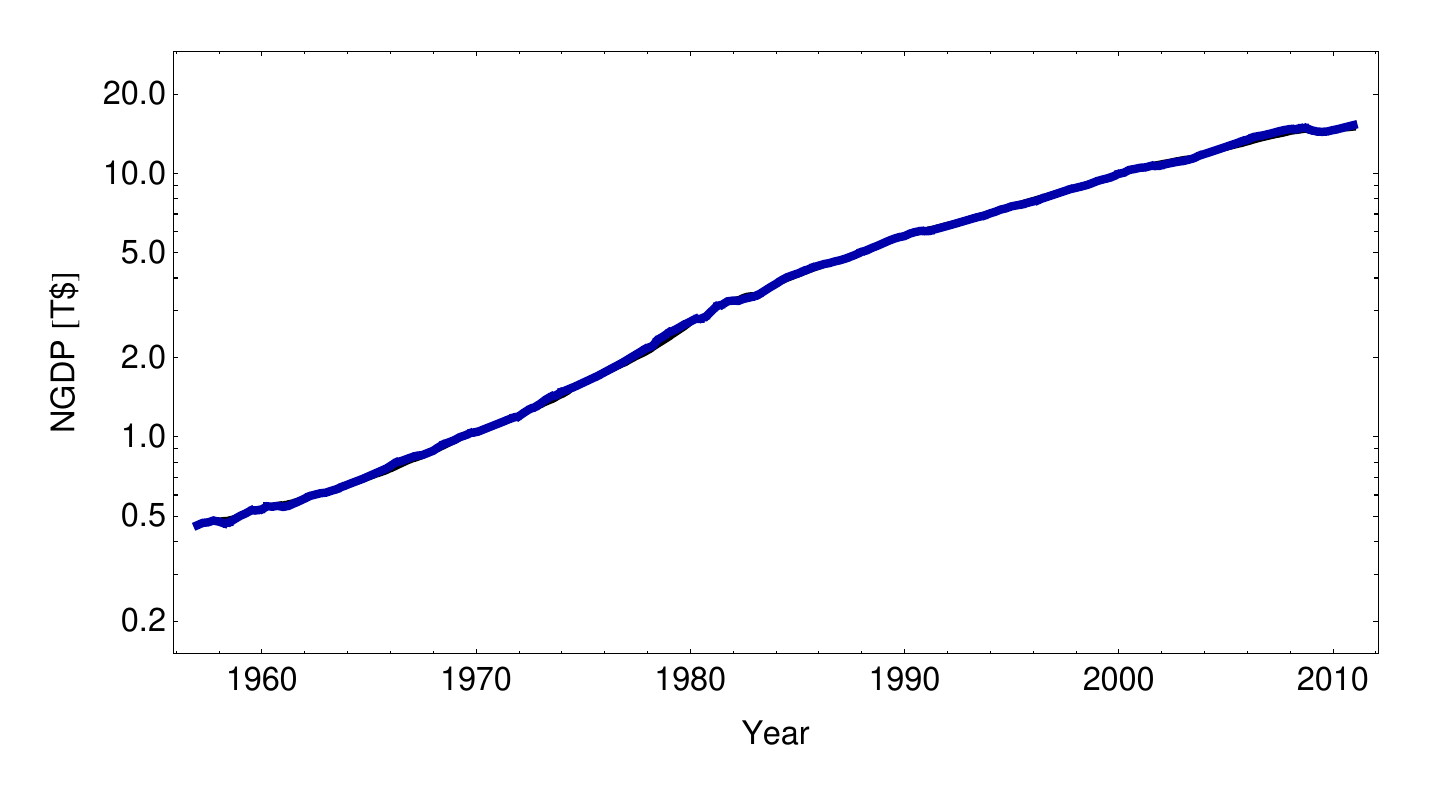}}\hfill{}\subfloat[Growth rate]{
\begin{centering}
\includegraphics[width=0.45\columnwidth]{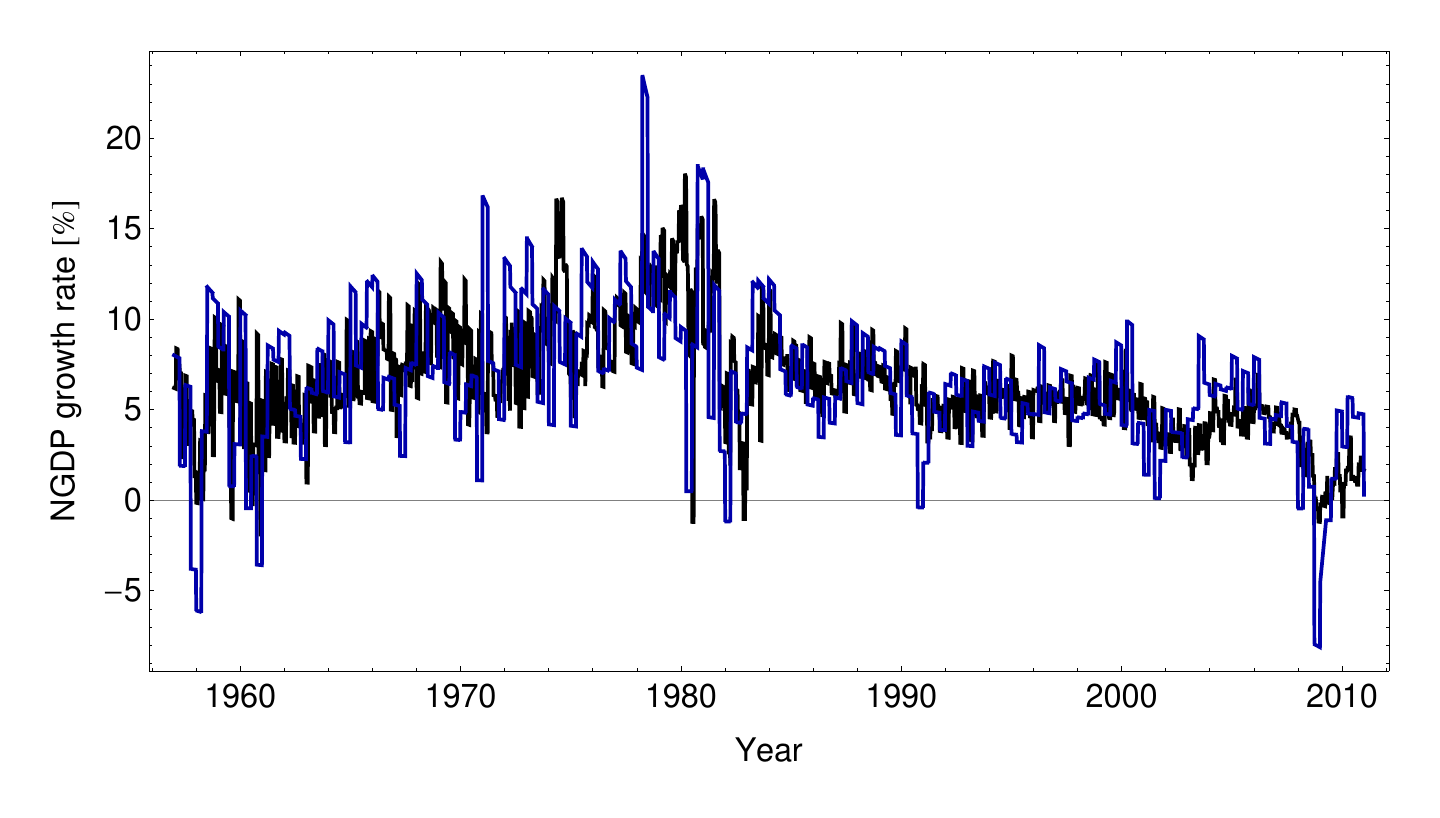}
\par\end{centering}
}\caption{\label{fig:solowsubfigures}Left: Nominal output using the Cobb-Douglas production function. Right: Growth rate of nominal output using the Cobb-Douglas production function.}
\end{figure}

Equation (\ref{eq:solow2}) is the generic Cobb-Douglas form. In the information equilibrium model, the exponents are free to take on any value (not restricted to constant returns to scale, i.e. $k_{1} + k_{2} = 1$). The resulting model is remarkably accurate as seen in Figure \ref{fig:solowsubfigures}.
It also has no changes in so-called total factor productivity ($A$ is constant). The results above use nominal capital and nominal GDP $N$ rather than the usual real capital and real output (RGDP, $R$). We use the \cite{fred} data series RKNANPUSA666NRUG for the real capital stock (capital stock at constant prices) and inflate to nominal capital stock via CPI less food and energy (CPILFESL).

\begin{figure}[t]
\centering{}\includegraphics[width=0.8\columnwidth]{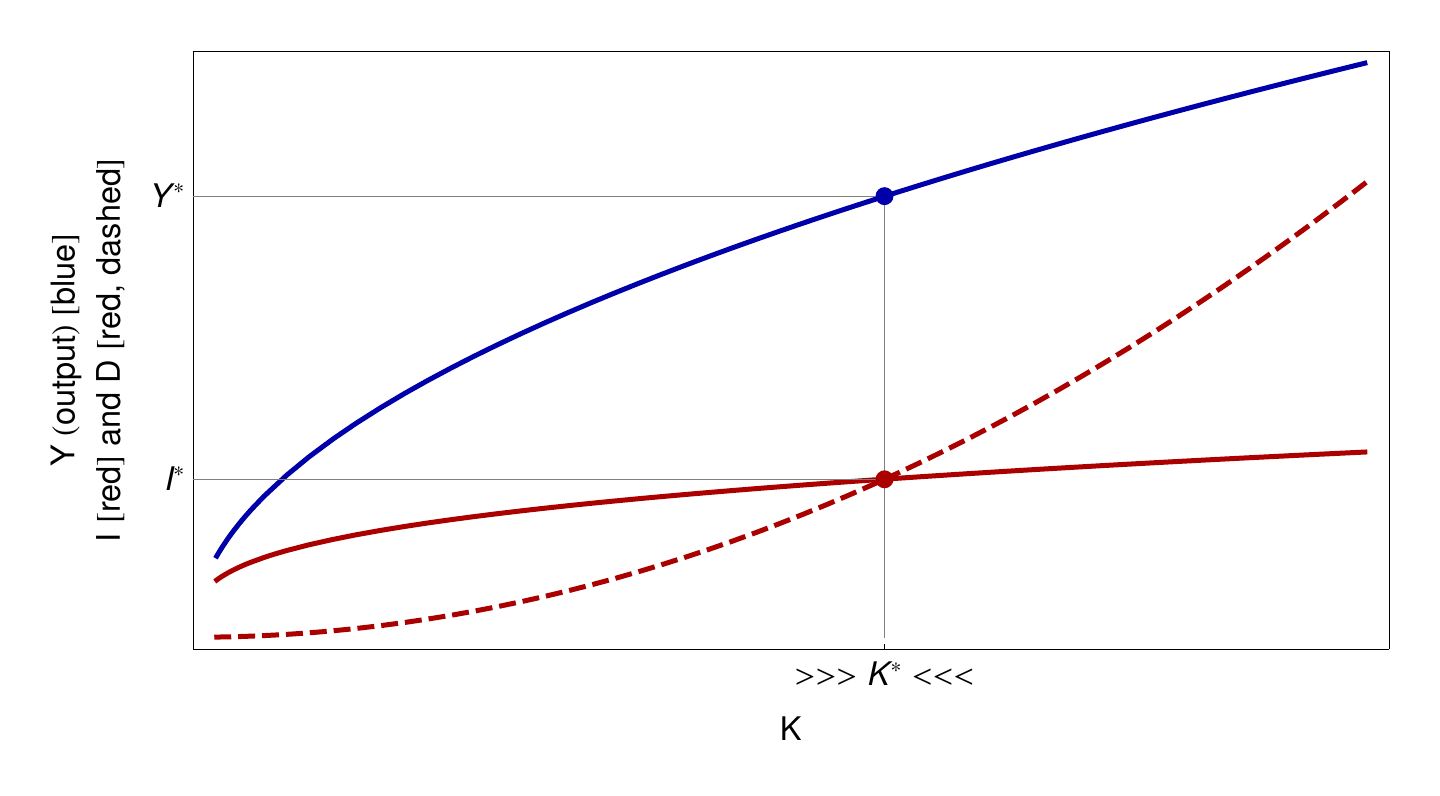}\caption{\label{fig:solowcapital}The Solow growth model as an information equilibrium model. Output is in blue, while investment is in red and depreciation is dashed red.}
\end{figure}

Let us assume two additional information equilibrium relationships with capital $K$ being the information source and investment $I$ and depreciation $D$ (include population growth in here if desired) being information destinations. In the notation we have been using: $K \rightleftarrows I$ and $K \rightleftarrows D$. This immediately leads to the solutions of the differential equation Eq.~(\ref{eq:main}):
\[
\frac{K}{K_{0}} = \left( \frac{D}{D_{0}}\right)^{\delta}
\]
\[
\frac{K}{K_{0}} = \left( \frac{I}{I_{0}}\right)^{\sigma}
\]
Therefore we have (the first relationship coming from the Cobb-Douglas production function)
\[
N \sim K^{\alpha}
\]
\[
I \sim K^{1/\sigma}
\]
\[
D \sim K^{1/\delta}
\]
If $\sigma = 1/\alpha$ and $\delta = 1$ we recover the original Solow model, but in general any $\sigma > \delta$ allows there to be an equilibrium. Figure \ref{fig:solowcapital} represents a generic plot of the relationships above.
Assuming the relationships $K \rightleftarrows I$ and $K \rightleftarrows D$ hold simultaneously gives us the equilibrium value of $K = K^{*}$:
\[
K^{*} = K_{0} \exp \left( \frac{\sigma \delta \log I_{0}/D_{0}}{\sigma - \delta} \right)
\]
This equilibrium value represents simultaneous information equilibrium in the two markets $K \rightleftarrows I$ and $K \rightleftarrows D$. Fluctuations in the value of capital $K$ away from $K = K^{*}$ will experience an entropic force to return $K^{*}$, so the equilibrium $K^{*}$ would be stable. Entropic forces will be discussed in more detail in Section \ref{ss:entropic}.

As a side note, the small $K$ region in Figure \ref{fig:solowcapital} does not appear because it is not a valid region of the model. The information equilibrium model is not valid for small values of $K$ (or any process variable). That allows one to choose parameters for investment and depreciation that could be e.g. greater than output for small $K$ -- a nonsense result in the traditional Solow model, but just an invalid region of the model in the information equilibrium framework. Another useful observation is that $N$ and $I$ have a supply and demand relationship in partial equilibrium with capital being demand and investment being supply since by transitivity (see Appendix \ref{app:equiv}) they are in information equilibrium (i.e. $N \rightleftarrows K$).

There might be more to the information equilibrium picture of the Solow model than just the basic mechanics -- in particular we might be able to analyze dynamics of the savings rate relative to demand shocks. We have built the model:
\[
N \rightleftarrows K \rightleftarrows I
\]
Where $N$ is output, $K$ is capital and $I$ is investment. Since information equilibrium is an equivalence relation (see Appendix \ref{app:equiv}), we have the model:
\[
p_{I}: N \rightleftarrows I
\]
with abstract price $p_{I}$. If we write down the differential equation resulting from that model
\begin{equation}\label{eq:outputinvestment}
p_{I} = \frac{dN}{dI} = \frac{1}{\eta} \; \frac{N}{I}
\end{equation}
There are a few things we can glean from this that are described below using general equilibrium, partial equilibrium, and making a connection to interest rates.

\subsubsection{General equilibrium in the Solow model}\label{sss:solowGE}

We can solve equation (\ref{eq:outputinvestment}) under general equilibrium giving us $N \sim I^{1/\eta}$. Empirically, we have $\eta \simeq 1$. Combining that with the results from the Solow model, we have
\[
N \sim K^{\alpha}
\]
\[
K \sim I^{\sigma}
\]
\[
N \sim I
\]
which tells us that $\alpha \simeq 1/\sigma$ -- one of the conditions that gave us the original Solow model result.

\subsubsection{Partial equilibrium in the Solow model}\label{sss:solowPE}

\noindent Since $N \rightleftarrows I$ we have a supply and demand relationship between output and investment in partial equilibrium. We can use equation (\ref{eq:outputinvestment}) and $\eta = 1$ to write
\[
I = (1/p_{I}) N \equiv s N
\]
where we have defined the saving rate as $s \equiv 1/(p_{I} \eta)$ to be (the inverse of) the abstract price $p_{I}$ in the investment market. A shock to aggregate demand would be associated in a fall in the abstract price and thus a rise in the savings rate. Overall, an economy does not always have pure supply or demand shocks, so there might be some deviations from a pure demand shock view. In particular, a "supply shock" (investment shock) should lead to a fall in the savings rate.

\subsubsection{Interest rates in the Solow model}\label{sss:solowinterest}

\noindent If we add the IS-LM model from section \ref{ss:islm} to include the interest rate ($i$) model using $N \rightleftarrows I \rightleftarrows M$ written in terms of investment and the money supply/base money:
\[
(i \rightleftarrows p) : I \rightleftarrows M
\]
where $p$ is the abstract price of money (which is in IE with the interest rate), we have a pretty complete model of economic growth that combines the Solow model with the IS-LM model. The interest rate model in Figure \ref{fig:longrate} joins the empirically accurate Cobb-Douglas production function in this section Figure \ref{fig:solowsubfigures}.

\subsection{A note on constructing models}

\begin{figure}[t]
\centering{}\includegraphics[width=0.8\columnwidth]{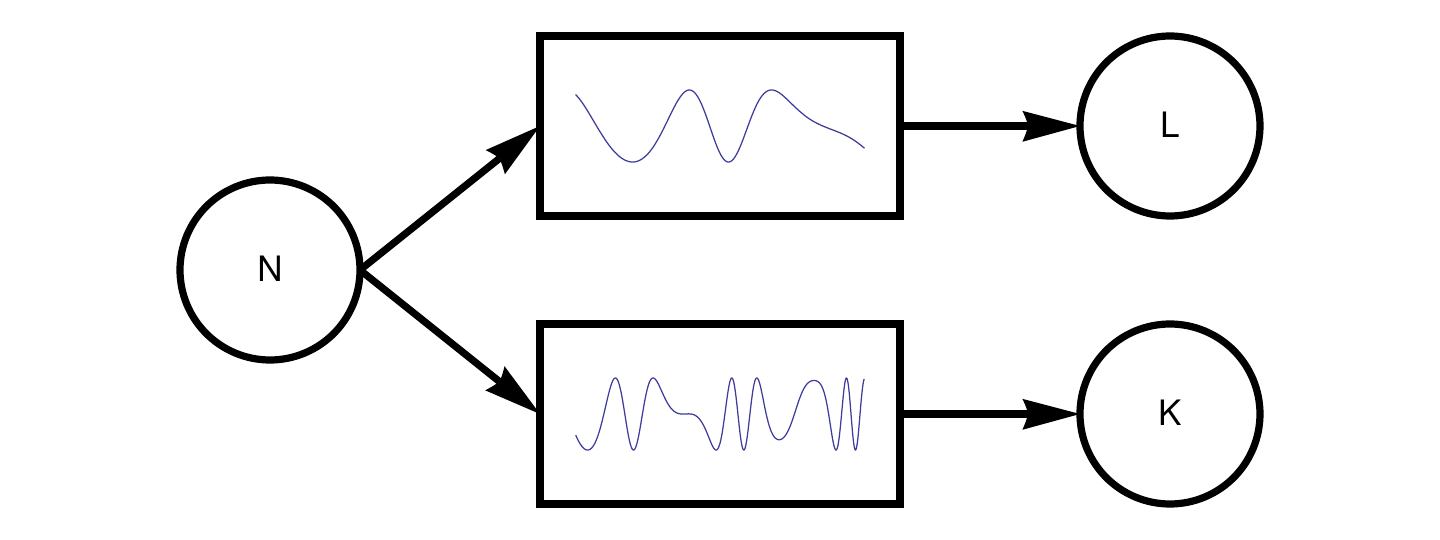}\caption{\label{fig:islminfodiagram}The Solow-Swan production functio as an information equilibrium model.}
\end{figure}

\begin{figure}[t]
\centering{}\includegraphics[width=0.8\columnwidth]{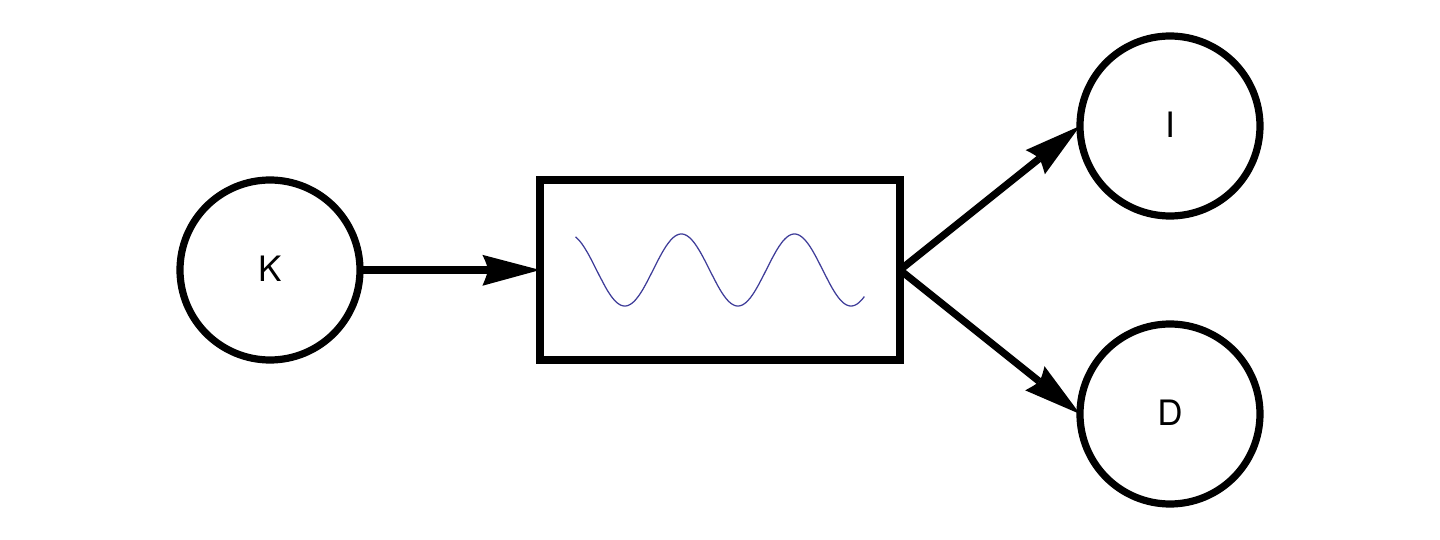}\caption{\label{fig:solowinfodiagram}The Solow growth model as an information equilibrium model.}
\end{figure}

\noindent In the previous sections we have used simultaneous markets that look formally the same. However, they are interpreted differently:
\begin{itemize}
\item In the Solow-Swan model, we used $N \rightleftarrows K$ and $N \rightleftarrows K$ to define the production function. These are taken to be \emph{independent equations} in general equilibrium. The represent two channels of information flow to two destinations as shown in Figure \ref{fig:solowinfodiagram}.
\item In the Solow-Swan model, we also used $K \rightleftarrows I$ and $K \rightleftarrows D$. These were taken to be \emph{simultaneous equations} in general equilibrium. The information transfer figure would look like a single channel with two destinations as shown in Figure \ref{fig:islminfodiagram}.
\item In the IS-LM model, we used $N \rightleftarrows M$ and $N \rightleftarrows S$. These are taken to be simultaneous equations in \emph{partial equilibrium} (i.e. $N$ moves slowly). The information transfer figure would look like the single channel Solow-Swan model diagram in Figure \ref{fig:islminfodiagram}. 
\end{itemize}

\section{Statistical economics}\label{s:statecon}

\noindent Analogies between physics and economics only have merit inasmuch as they are useful. In this section we will take some initial steps toward defining the ``statistical economics'' of \cite{Smolin:2009} analogous to statistical mechanics. Consider a collection of individual market information sources $\{N_{i}\}$. In the following we will work in ``natural units'' and take $n_{i} = N_{i}/N_{i, ref}$ and $m = M/M_{ref}$. The $n_{i}$ are the demands in the individual markets and $m$ is the money supply (it does not matter which aggregate at this point). The individual markets are the solutions to the equations:
\begin{equation}\label{eq:submarkets}
\frac{d n_{i}}{d m} = a_{i} \frac{n_{i}}{m}
\end{equation}
following from the introduction of the money-mediated information transfer model $n_{i} \rightleftarrows m \rightleftarrows s_{i}$ as was shown in Section \ref{ss:adasmodel}. One interesting thing is that the defining quality of these individual markets -- equation (\ref{eq:submarkets}) leads to supply and demand diagrams -- is homogeneity of degree zero in the supply and demand functions (as noted in Section \ref{ss:alt}), which is one of the few properties that survive aggregation in the Sonnenschein-Mantel-Debreu theorem.

Now consider the sum (defining aggregate nominal output or NGDP across all the markets)
\begin{equation}
N(m) =\sum_{i} n_{i} =  \sum_{i} m^{a_{i}} = \sum_{i} e^{a_{i} \log m}
\end{equation}
This has a form similar to a partition function\footnote{Partition functions represent maximum entropy probability distributions; the mathematical formalism is similar to random utility discrete choice models.}
\begin{equation}
Z(\beta) =  \sum_{i} e^{-\beta E_{i}}
\end{equation}
Proceeding by analogy, we will define the macroeconomic partition function to be:
\begin{equation}
Z(m) \equiv \sum_{i} \frac{1}{n_{i}} =  \sum_{i} m^{-a_{i}} = \sum_{i} e^{-a_{i} \log m}
\end{equation}
With this partition function, the ensemble average (or expectation value, denoted with angle brackets) of the exponent $a_{i}$ is:
\begin{equation}
\label{eq:apart}
\langle a \rangle = -\frac{\partial \log Z(m)}{\partial \log m} = \frac{\sum_{i} a_{i}e^{-a_{i} \log m}}{\sum_{i} e^{-a_{i} \log m}}
\end{equation}
which corresponds to the aggregate information transfer index $k = \langle a \rangle$. Additionally, the nominal economy will be the number of markets $N_{0}$ times the ensemble average of an individual market $m^{a_{i}}$, i.e.
\begin{eqnarray}
\langle N(m) \rangle & = & N_{0} \frac{\sum_{i} m^{a_{i}} e^{-a_{i} \log m}}{\sum_{i} e^{-a_{i} \log m}}\\
& = & N_{0} \frac{\sum_{i} 1}{\sum_{i} e^{-a_{i} \log m}} = \frac{N_{0}^{2}}{Z(m)}\label{eq:npart}
\end{eqnarray}
Equation (\ref{eq:npart}) simplifies when $M = M_{ref}$ ($m = 1$) to
\[
\langle N(1) \rangle = N_{0}^{2}/N_{0} = N_{0}
\]
First there is an interesting new analogy with thermodynamics: $\log m$ is playing the role of $\beta = 1/kT$, the Lagrange multiplier (thermodynamic temperature). As $m$ gets larger the states with higher $a_{i}$ (high growth markets) become less probable, meaning that a large economy (with a large money supply) is more like a cold thermodynamic system. The meaning of large here is measured by $M_{ref}$. As an economy grows, it cools, which leads to slower growth -- going by the terms "the great stagnation" in \cite{Cowen:2011} or "secular stagnation" in \cite{Summers:2013} -- and as we shall see a bending of the price level vs money curve (low inflation in economies with large money supplies). 

Let us take $N_{0} =$ 100 random markets with normally distributed $a_{i}$ with average $\bar{a} = 1.5$ and standard deviation $\sigma_{a} = 0.5$ and plot 500 Monte Carlo runs of the information transfer index $\langle a \rangle$, the price level $\langle a m^{a -1}\rangle$ and the nominal output $\langle N(m) \rangle$.
\begin{figure}[t]
\begin{centering}
\includegraphics[width=0.7\columnwidth]{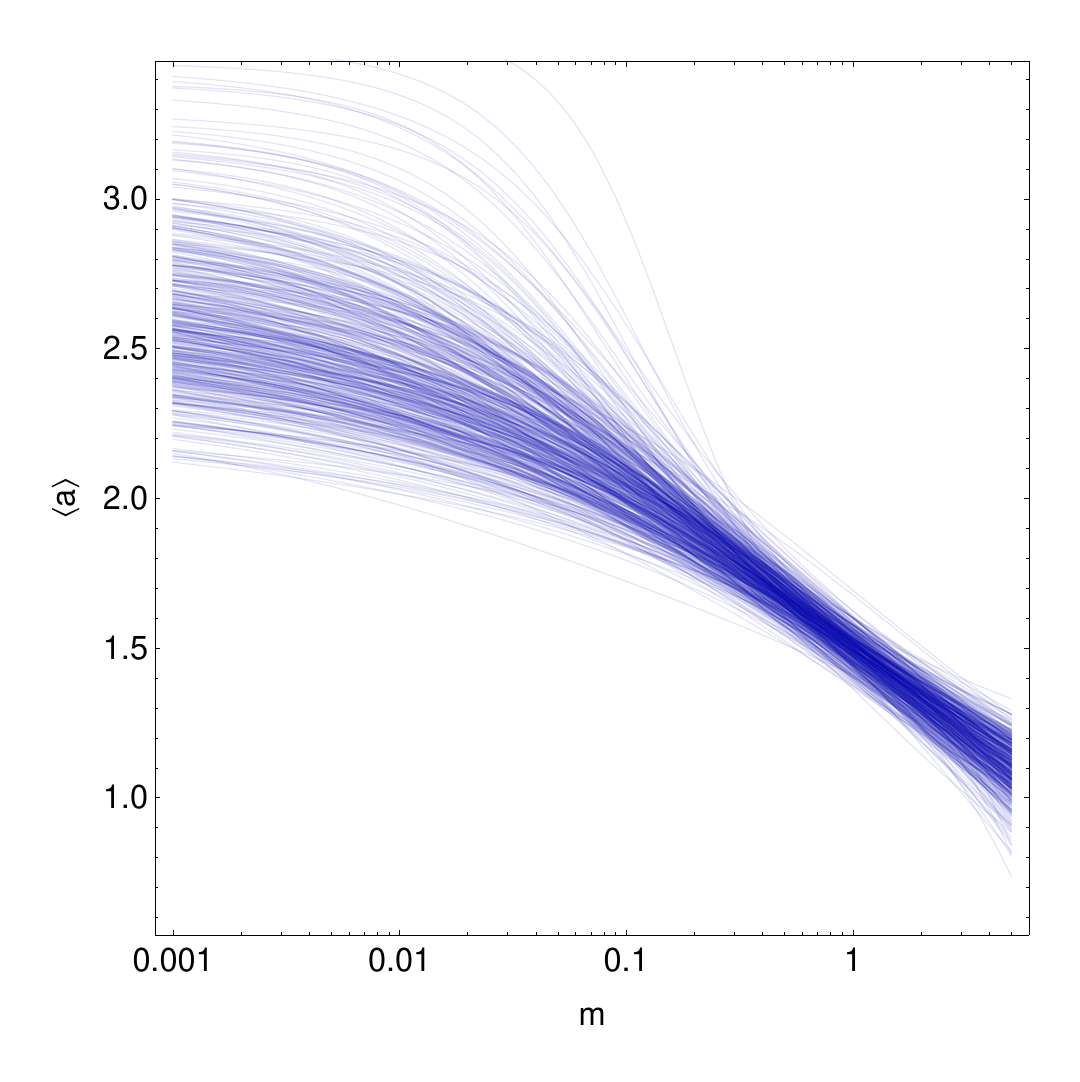}
\par\end{centering}
\caption{\label{fig:partition1}Partition function calculation of the ensemble average of the information transfer index from Eq.~(\ref{eq:apart}).}
\end{figure}
\begin{figure}
\begin{centering}
\includegraphics[width=0.7\columnwidth]{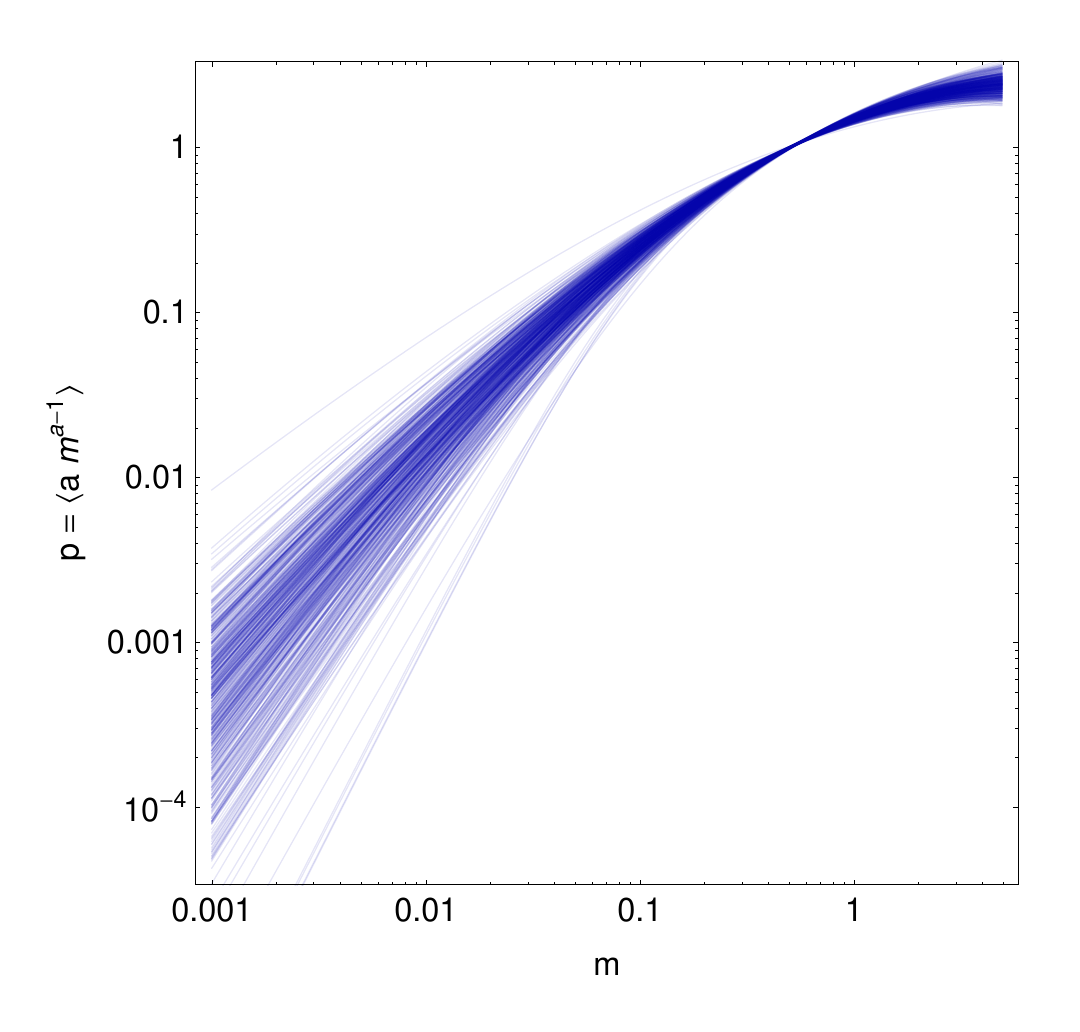}
\par\end{centering}
\caption{\label{fig:partition2}Partition function calculation of the ensemble average of the price level $\langle a m^{a - 1} \rangle$.}
\end{figure}
\begin{figure}
\begin{centering}
\includegraphics[width=0.7\columnwidth]{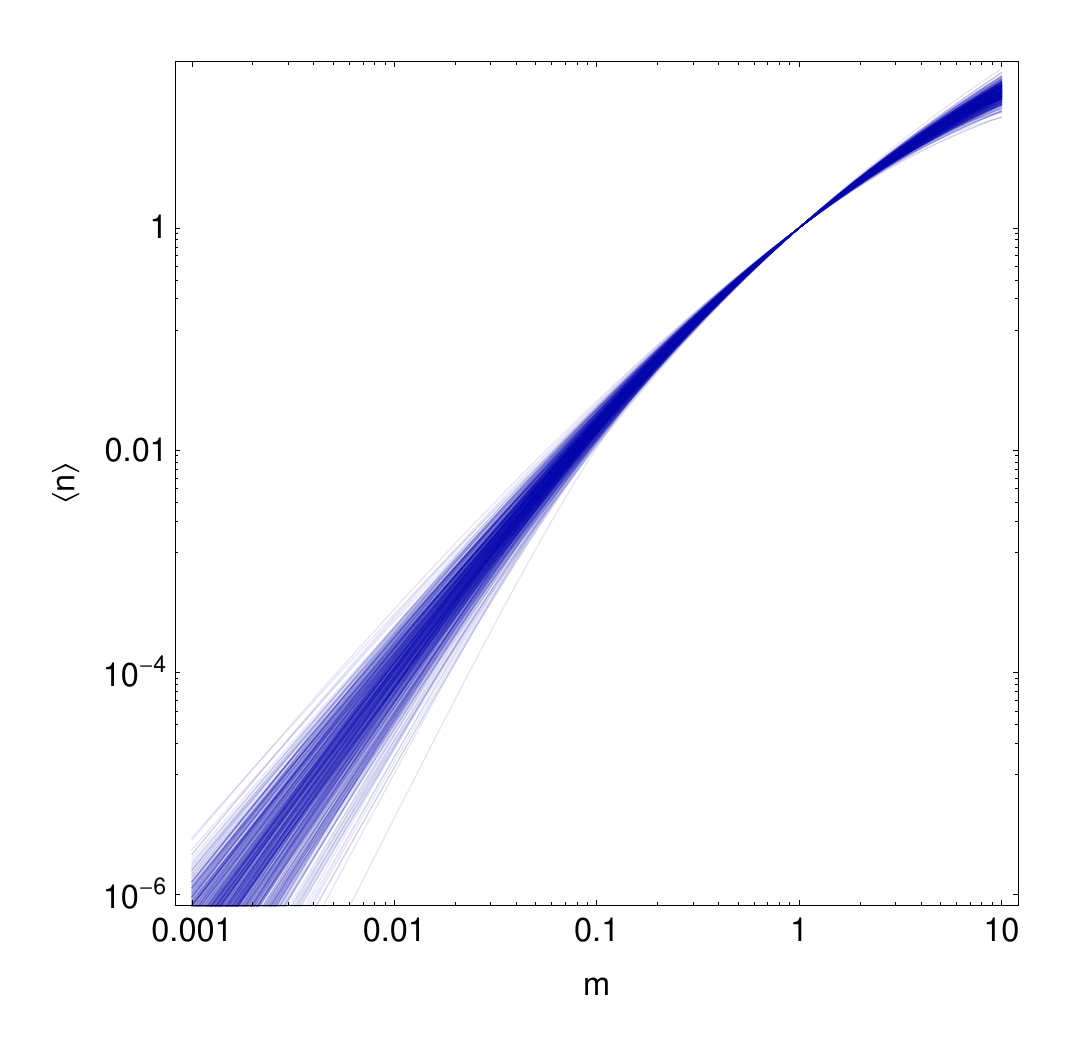}
\par\end{centering}
\caption{\label{fig:partition3}Partition function calculation of the ensemble average of nominal output from Eq.~(\ref{eq:npart}).}
\end{figure}
In Figure \ref{fig:partition1} we can see the economies start out well described by the quantity theory ($k \approx 2$) and move towards lower $\langle a \rangle$ as the money supply increases. We can see the bending of the price level versus money supply in Figure \ref{fig:partition2}. In Figure \ref{fig:partition3}, we can see the trend towards lower growth relative to the growth in the money supply.

\begin{figure}[t]
\centering{}\includegraphics[width=0.8\columnwidth]{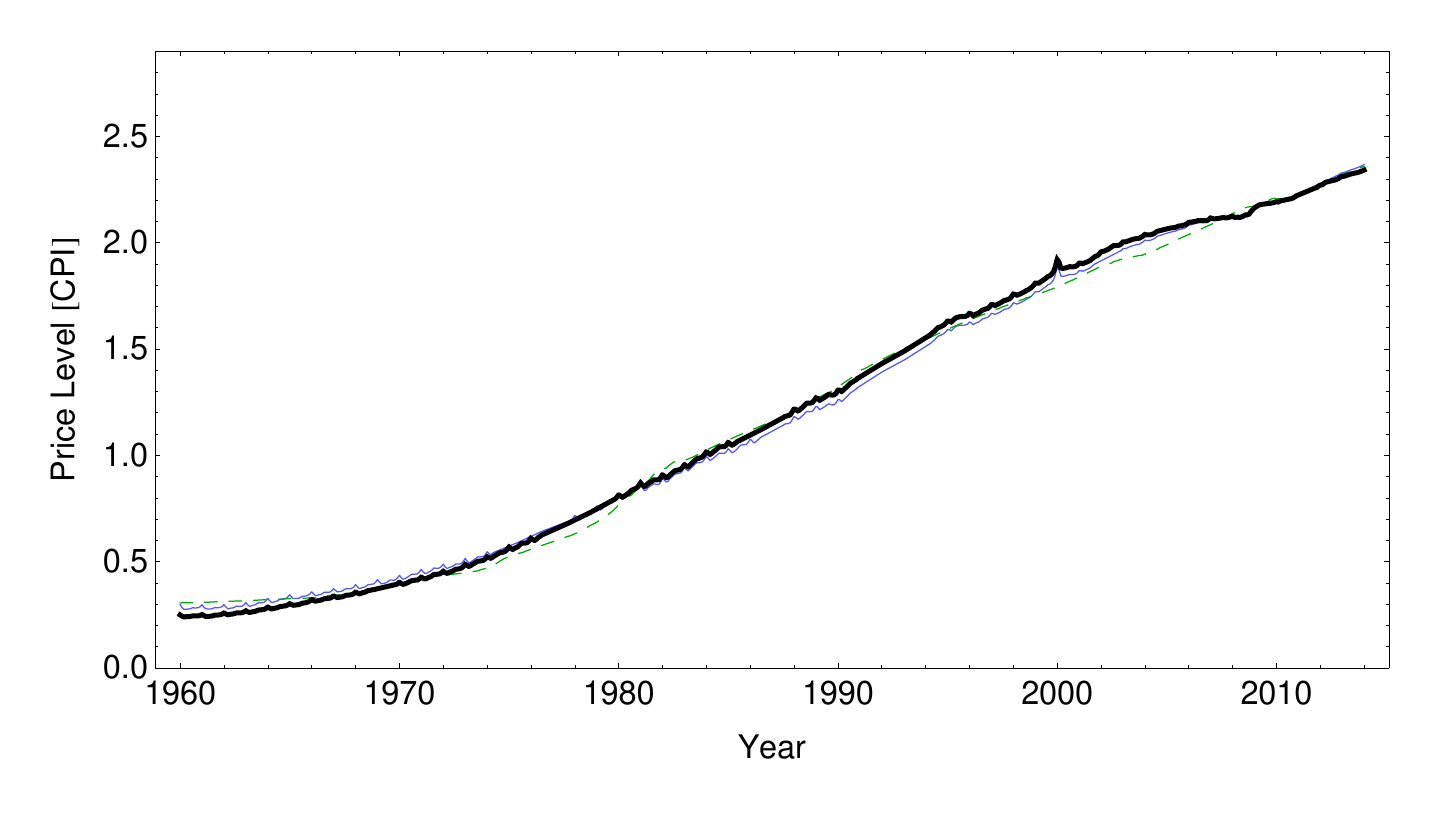}\caption{\label{fig:partplevelandQTM}Model of price level from the partition function approach and the approach of Section \ref{ss:qtm} using core CPI data from \cite{fred}. The deviations are small between the models.}
\end{figure}

\begin{figure}[t]
\centering{}\includegraphics[width=0.8\columnwidth]{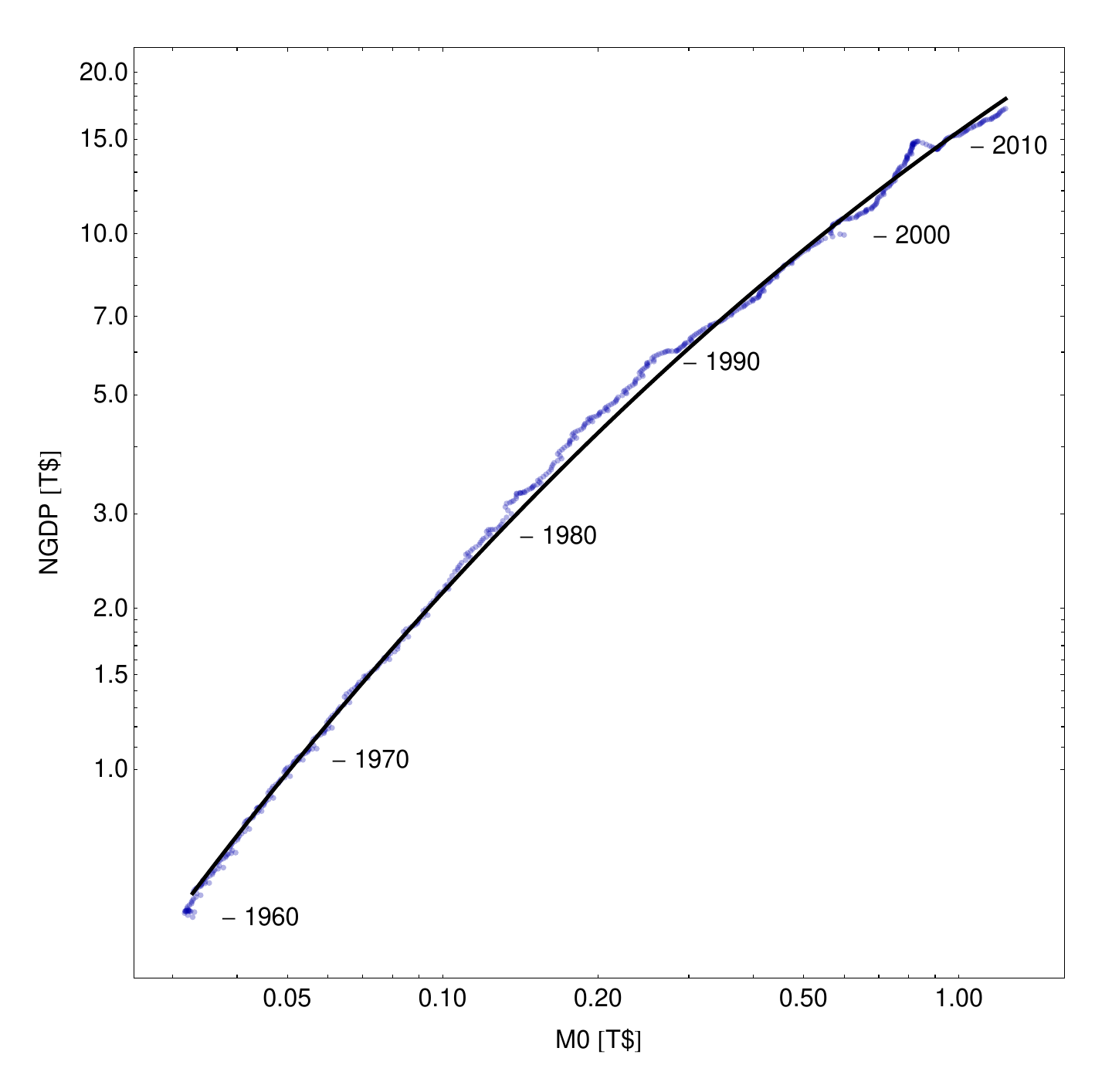}\caption{\label{fig:pathngdp}Model of nominal output.}
\end{figure}

The question now is: how well does this oversimplified picture work with real data? After normalizing the price level and scaling the money supply, the function $P = \langle a m^{a-1} \rangle$ almost exactly matches the information transfer model for the price level in Section \ref{ss:qtm}. The information transfer model of Section \ref{ss:qtm} and the partition function version above are graphed in Figure \ref{fig:partplevelandQTM}. There are only small deviations.
We apply the ensemble average result calculated using Eq.~(\ref{eq:npart}) and presented in Figure \ref{fig:partition3} to empirical data for the US and show it in Figure \ref{fig:pathngdp}. This general trend is frequently encountered in the data for several countries as part of a growing survey, see \cite{Smith:2015a}, and will be explored in future work.

With the partition function approach, we can see that reduced inflation with a large money supply (a thermodynamically colder system) as well as reduced growth in Figure \ref{fig:pathngdp} are \emph{emergent} properties. They do not exist for the individual markets; it is important to emphasize this aspect of the model. An economy with a larger money supply is more likely to be realized as a large number of lower growth states (higher entropy) than a smaller number of high growth states\footnote{A physical example: a state consisting of many low energy photons has higher entropy than a state of equal energy consisting of a few high energy photons. Therefore the blackbody radiation spectrum tends to be produced by a distribution consisting of more photons with lower energy than one with a few photons of higher energy.}. One can think of the exponents $a_{i}$ as growth states where the available $a_{i}$ depend macroeconomic conditions.

\subsection[Entropic forces]{Entropic forces and emergent properties}\label{ss:entropic}

\noindent There are several novel interpretations of observed or theorized macroeconomic effects that come from this partition function treatment. First, partition functions are maximum entropy distributions so macroeconomic equilibrium may be thought of as a maximum entropy state. Second, while the trend towards lower $\langle a \rangle$ is apparent in the ensemble average, there is no microeconomic rationale. Lower growth as an economy increases in size is an ``emergent'' property of economies. Larger economies are ``cold'', but small economies without asymptotically large $N_{0}$ do not have a well-defined ``temperature'' and individual markets may dominate output.

An additional emergent property is the slow decline in the response of the price level to changes in the monetary base since
\begin{equation}
\log P \sim (\langle a \rangle - 1) \log M
\end{equation}
Using the interest rate model of section \ref{ss:islm} to connect $M$ to $i$ means that the impact of monetary policy is reduced for larger, ``colder'' economies. Lowering interest rates (expanding the base) has a smaller and smaller effect as $\langle a \rangle$ falls. This idea of ineffective monetary policy is similar to the concept of the liquidity trap, see e.g. \cite{Krugman:1998}, however there are some key differences:
\begin{itemize}
\item This \emph{information trap} does not depend on the zero lower bound for interest rates. However, lower interest rates are related. As $\langle a \rangle \rightarrow 1$, $\partial P/\partial M \rightarrow 0$ so that if $i^{k_{i}} \sim N/M = PY/M$, increasing $M$ will lower interest rates (this describes the liquidity effect). Therefore $\langle a \rangle \approx 1$ will tend to be associated with lower interest rates.
\item The information trap does not have a sudden onset, but is part of a gradual trend towards lower interest rates. The onset may appear sudden in economies that use monetary policy for macroeconomic stabilization when a large shock hits (for example, the global financial crisis of 2008) and monetary policy appears more ineffective than during previous shocks.
\item There is no microeconomic rationale for the information trap. One mechanism for the liquidity trap is based on the idea that at the zero lower bound for interest rates, there is no difference between short term treasury securities and zero-interest money. In the information equilibrium approach, the information trap is an emergent property dependent on an ensemble of markets.
\end{itemize}
Finally, interpreting the $a_{i}$ realized in an ensemble of markets as the occupied growth states in an economy lets us form a novel hypothesis: nominal rigidity is an entropic force. Entropic forces in thermodynamics are forces that have no microscopic analogy, yet have observable macroscopic effects. One of the most commonly encountered physical entropic forces is diffusion\footnote{Gravity may be an entropic force as well, and if it is true would actually be the best example. See \cite{Verlinde:2011}.}. If molecules are initially distributed on one side of a container, in short order they become uniformly distributed throughout the container. There is no microscopic force on a molecule proportional to local deviations in density $\rho(r)$ from average density $\rho_{0}$ (the volume is $V$)
\[
F(r) \sim \rho(r) - \rho_{0}
\]
\[
\rho_{0} \equiv \frac{1}{V} \int d^{3}r \rho (r)
\]
However molecules behave \emph{in aggregate} as if such a microscopic force existed, evening the distribution of molecules and producing a uniform distribution. Individual molecules feel no such force. If the distribution is perturbed away from a uniform distribution, it will feel an entropic force to return to the original uniform distribution. Thus in thermodynamics we say there exists an entropic force (diffusion) maintaining a uniform distribution.

Returning to our ensemble of markets we can imagine an equilibrium distribution of growth states $a_{i}$. Analogous to the physical system, markets will feel an entropic force to maintain the distribution of growth states set by macroeconomic observables. Growth states will not spontaneously over-represent the negative (or simply sub-inflation) growth states and will behave as if there was a force keeping them in the distribution. Specifically, while the distribution of prices (or wages) in the economy may not adjust to adverse shocks as an aggregate (e.g. the price level will not fall), individual prices may fluctuate by a large amount, for example see \cite{Eichenbaum:2008}. Microfounded mechanisms like Calvo pricing (e.g. menu costs) enforcing nominal price or wage rigidity would be analogous to the fictitious density dependent force mentioned above for diffusion.

The physical concept of entropic forces is similar to the economic concept of \emph{t\^{a}tonnement}. In the case of entropic forces, individual molecules restore equilibrium by random chance because equilibrium is the most likely state. The molecules are ``coordinated'' by entropy. In the case of t\^{a}tonnement, individual agents restore equilibrium by announcing their guesses at equilibrium prices to the Walrasian auctioneer, who coordinates agent prices until equilibrium (zero excess demand/supply) is achieved. \cite{Jaynes:1991} referred to this process as ``dither'' and noted its relevance for economics.

We can take this entropic description further by analogy with physical systems. In the beginning of Section \ref{s:statecon} we connected $\log M$ (where M is the currency supply) with $\beta = 1 / k_{B}T$ in the partition function. If we take the thermodynamic definition of temperature:
\[
\frac{1}{k_{B} T} = \frac{dS}{dE}
\]
as an analogy (where $S$ is entropy and $E$ is energy), we can write
\begin{equation}
\log \frac{M}{\gamma M_{0}} = \frac{dS_{e}}{dN}
\end{equation}
where we have used the correspondence\footnote{In the model of the ideal gas in \cite{Fielitz:2014}, the information source is the energy of the system. In our case that is aggregate demand $N$.} of the demand (NGDP, or $N$) with the energy of the system. We do not assume what the economic entropy $S_{e}$ is at this point. However, if we take
\[
N \sim M^{k}
\]
then we can write down
\[
\frac{1}{k} \log \frac{N}{\gamma M_{0}} = \frac{dS_{e}}{dN}
\]
So that, integrating both sides (with $k$ being a slowly varying function of $N$), we obtain
\[
S_{e} = \frac{1}{k} \; \frac{N}{\gamma M_{0}} (\log \frac{N}{\gamma M_{0}} - 1) + C
\]
Using Stirling's approximation for large $N/\gamma M_{0} \gg 1$ allows us to write (dropping the integration constant $C$)
\begin{equation}
S_{e} \simeq \frac{1}{k} \log \; \left( \frac{N}{\gamma M_{0}}\right)!
\end{equation}
If we compare this equation with the Boltzmann definition of entropy
\[
S = k_{B} \log W
\]
We can identify $(N/(\gamma M_{0}))!$ with the number of microstates in the economy and $1/k$ being the `economic Boltzmann constant'. The factorial $N!$ counts the number of permutations of $N$ objects and one possible interpretation is that $\gamma M_{0}$ adjusts for the distinguishability of given permutations -- all the permutations where dollars are moved around in the same firm or industry are likely indistinguishable or approximately so. This could lend itself to an interpretation of the constant $\gamma$ across countries discussed in Appendix \ref{app:sigmakappa}: large economies are diverse and likely have similar relative sizes of their manufacturing sectors and service sectors, for example. Once you set the scale of the money supply $M_{0}$, the relative industry sizes (approximately the same in advanced economies) are set by $\gamma$. This picture provides the analogy that a larger economy ($N$) has larger entropy (economic growth produces entropy) and lower temperature ($1/\log M$).

\begin{figure}[t]
\centering{}\includegraphics[width=0.8\columnwidth]{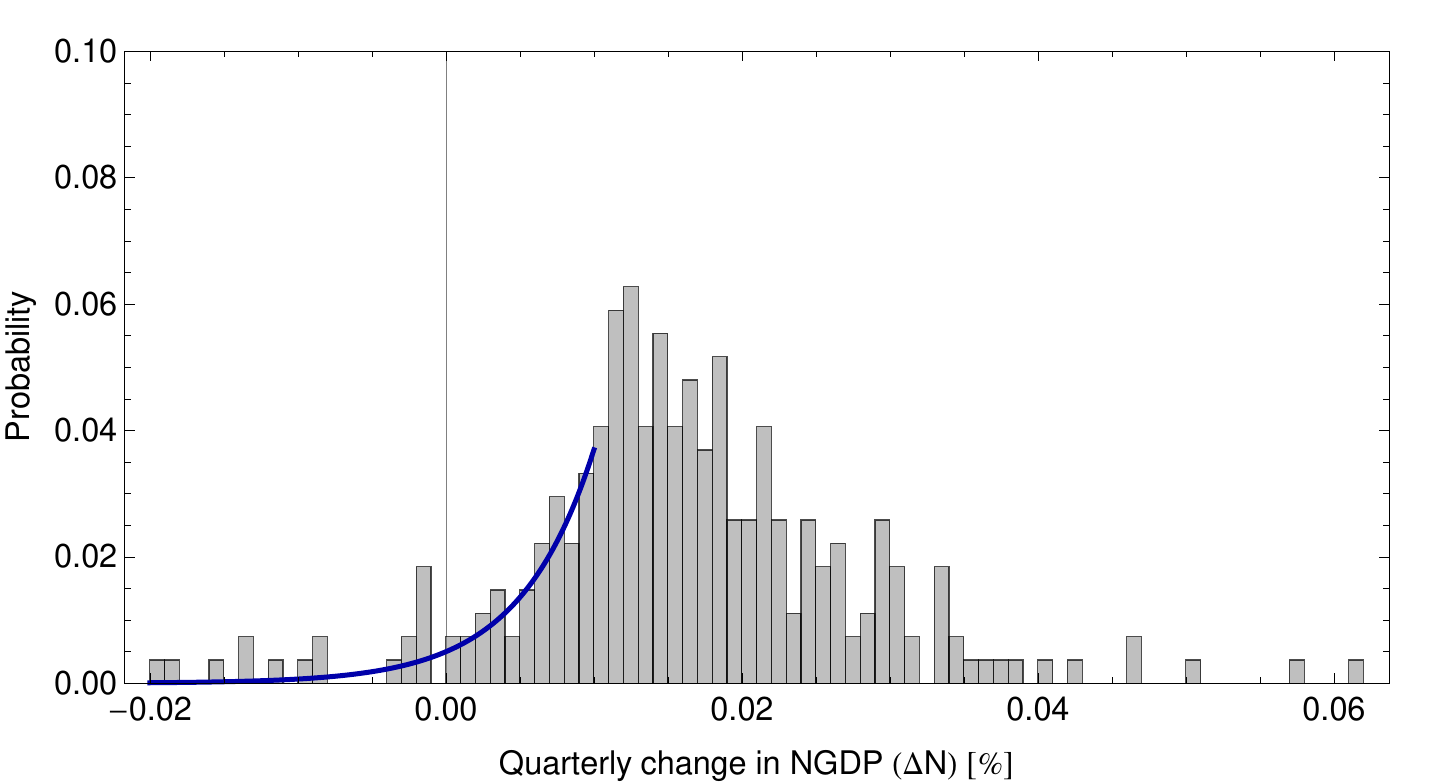}\caption{\label{fig:fluc}The distribution of quarterly changes in NGDP 1947-2015 (gray bars). Data from \cite{fred} series GDP. Heuristic estimate of the probability tail from an application of the fluctuation theorem is shown as a blue line.}
\end{figure}

For small changes in $N \rightarrow N+\Delta N$, we can show
\begin{equation}
\Delta S_{e} \simeq \frac{\Delta N}{k \gamma M_{0}} \log \frac{N}{\gamma M_{0}}
\end{equation}
Economic growth represents a rise in economic entropy. If the second law of thermodynamics applied to economic systems, then one would expect that $\Delta N > 0$. However this is not true in real macroeconomic systems. In particular, one heuristic indicator for a recession is two consecutive quarters of falling NGDP. The second law of thermodynamics is statistically violated on small scales per the so-called fluctuation theorem, see e.g.~\cite{Evans:2002}, however this would imply a specific form of the violation in terms of the probabilities $P$
\[
\frac{P(+\Delta N)}{P(-\Delta N)} = e^{\Delta N}
\]
The tail of the actual distribution of changes in NGDP is over-represented relative to a naive application\footnote{This is not intended as a rigorous argument, but rather simply to motivate the idea that falling NGDP in a recession is not a random event.} of this theoretical distribution as can be seen in Figure \ref{fig:fluc}. This is not a new observation; the fact that the distribution of changes in NGDP (and other markets) does not have exponential tails is a stylized fact of macroeconomics.

However there is another way an economic system could violate the second law of thermodynamics that is not available to a physical system composed of molecules: coordination among the constituents. An ideal gas that changes from a state where molecules have randomly oriented velocities to a state where velocities are aligned represents a large fall in the entropy of that ideal gas. This will not spontaneously happen with meaningful probability in large physical systems. In economic systems, agents will occasionally coordinate (for example, so-called ``herd behavior''), and this may be the source of the fall in economic entropy -- and hence output -- associated with recessions. It is also extremely unlikely that economic agents will re-coordinate themselves in order to undo the fall in NGDP. Absent reactions from the central bank or central government (monetary or fiscal stimulus), the return to NGDP growth will continue at the previous growth rate.

\section{Summary and conclusion}\label{s:summary}

\begin{figure}[t]\centering
\subfloat[Cobb-Douglas function for Mexico]{\centering{}\includegraphics[width=0.45\columnwidth]{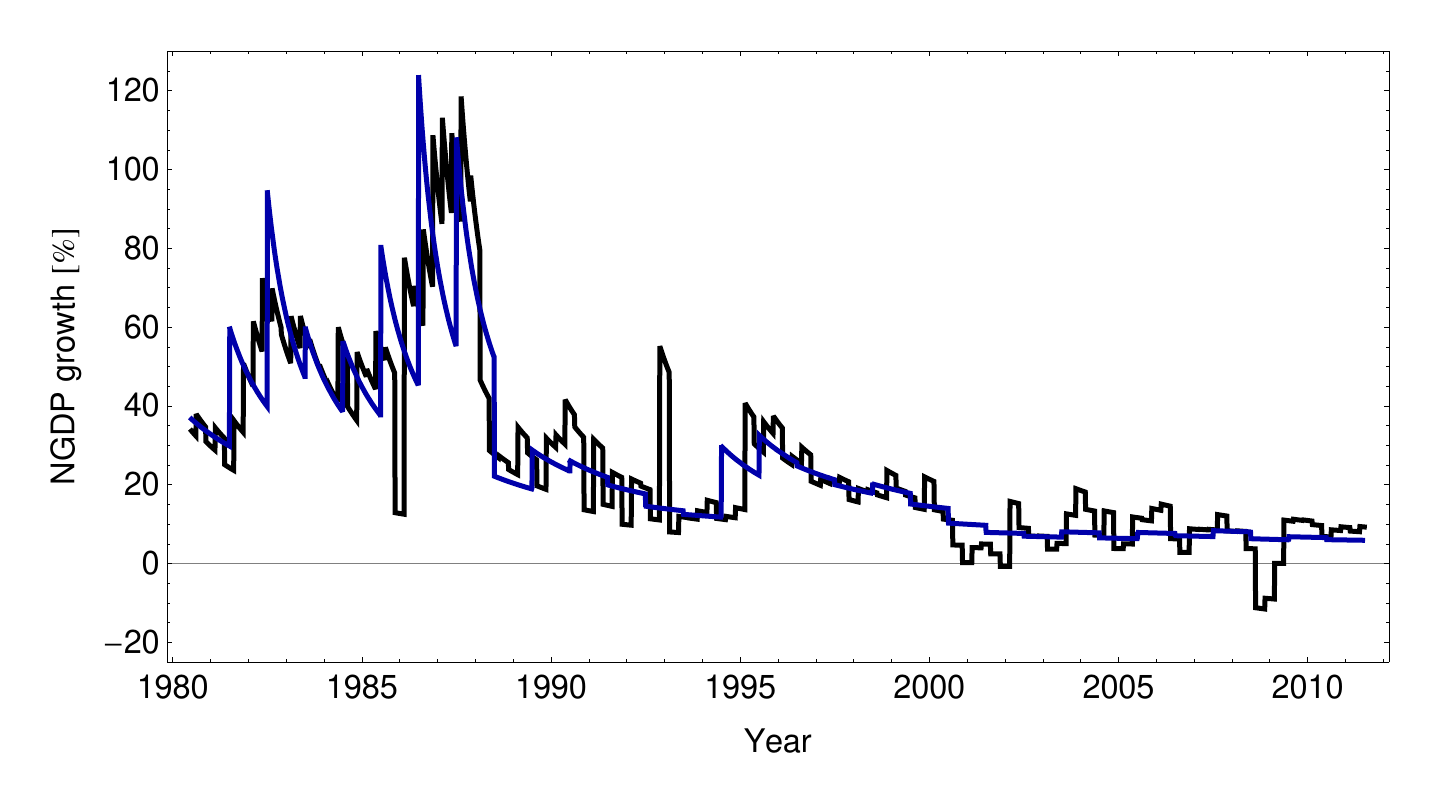}}\subfloat[Price level for Japan]{
\includegraphics[width=0.45\columnwidth]{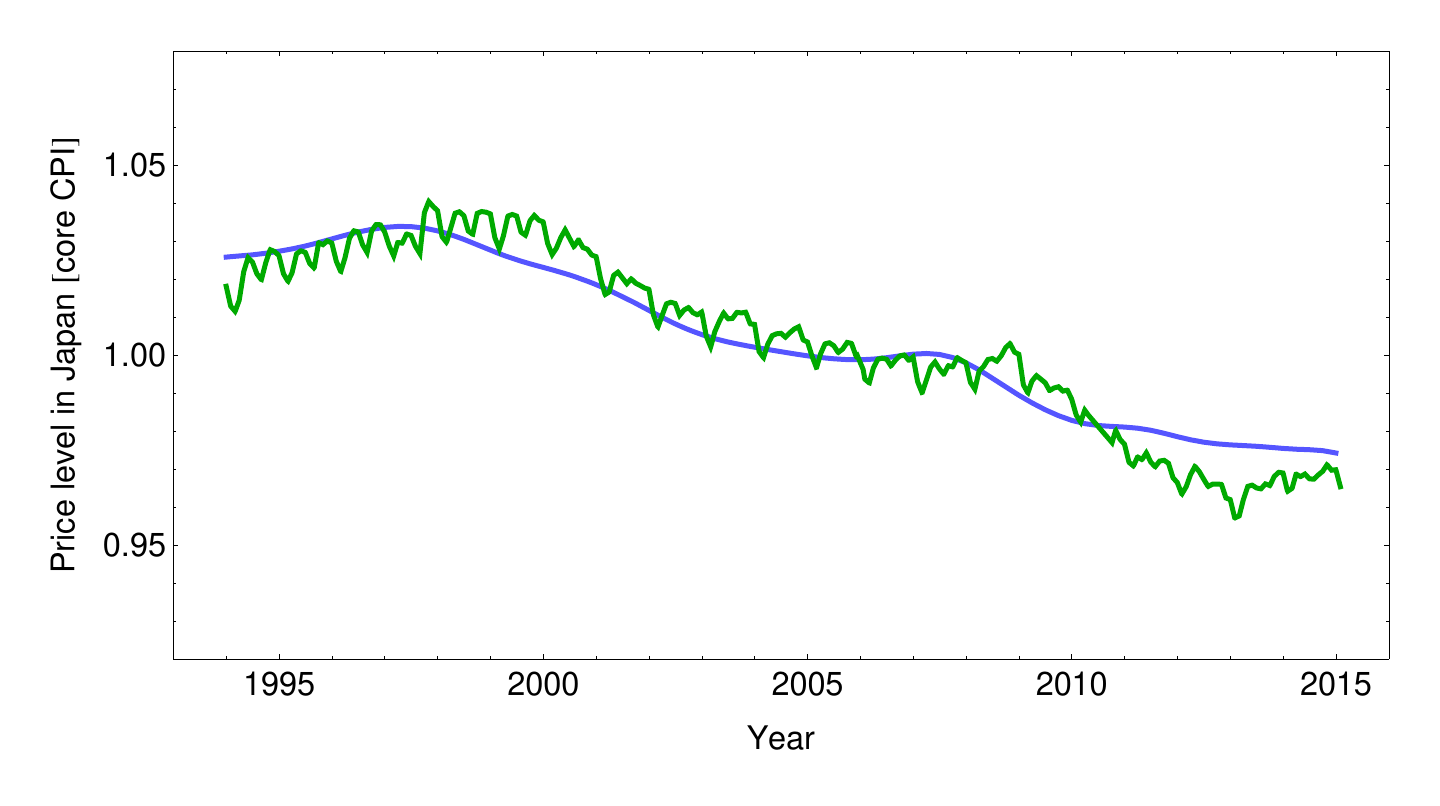}
\par\hfill{}}\par
\subfloat[Inflation rate for the EU]{\centering{}\includegraphics[width=0.45\columnwidth]{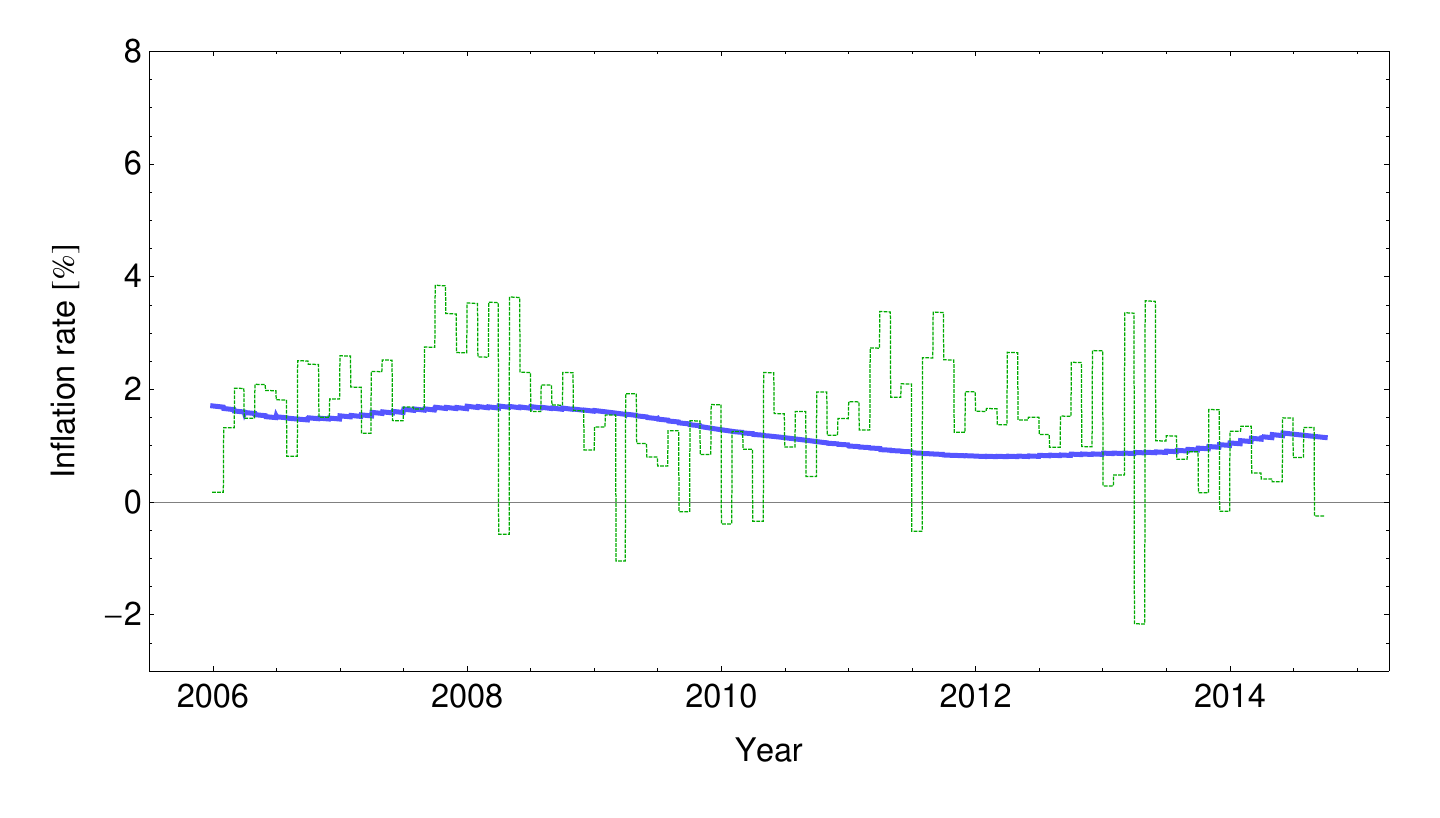}}\subfloat[Interest rates for the UK]{
\includegraphics[width=0.45\columnwidth]{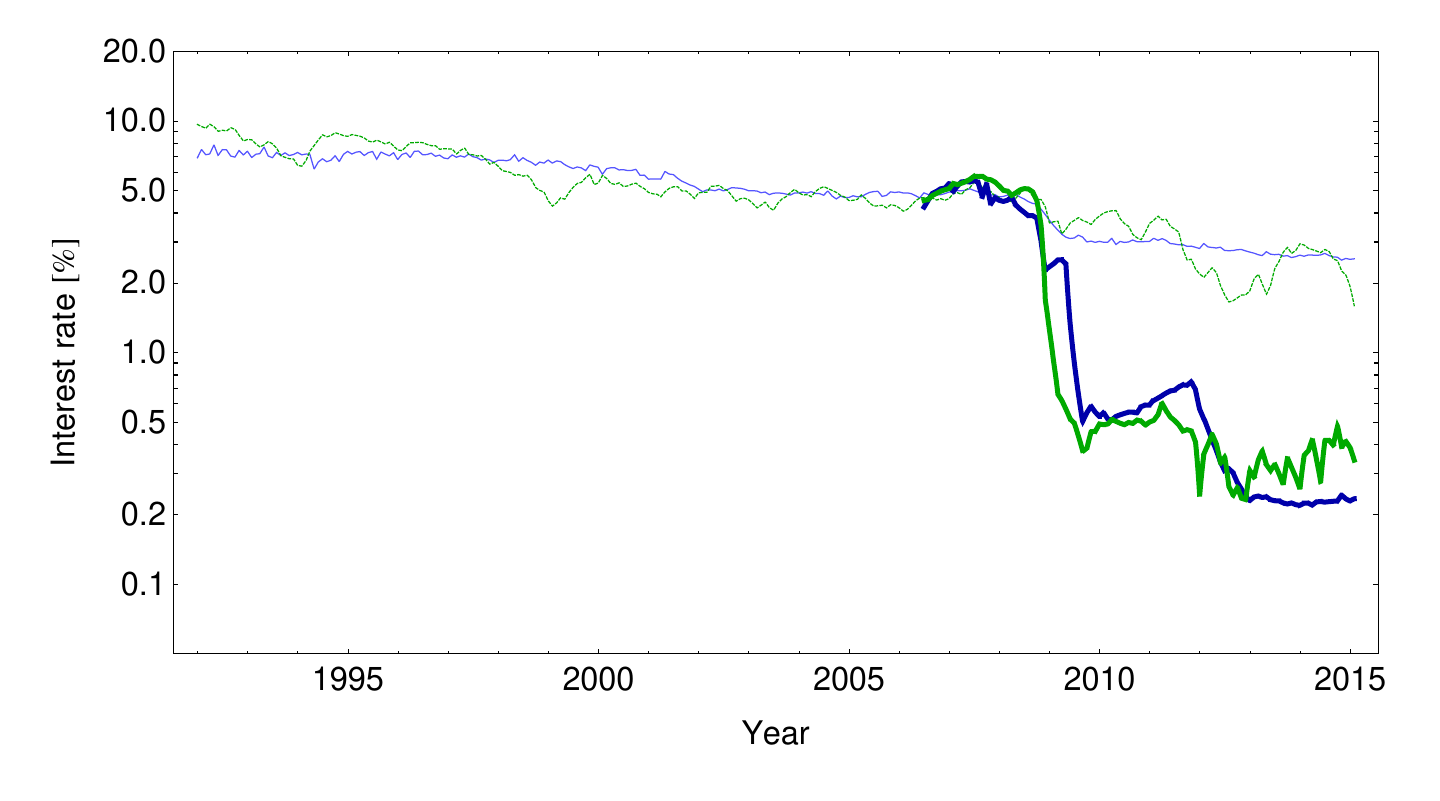}
\par}
\caption{\label{fig:othersubfigures}Application of information equilibrium to other countries. Nominal growth from the Cobb-Douglas production function (Mexico) in the Solow model, price level (Japan), inflation rate (EU) and long- and short-term interest rates (UK).}
\end{figure}

\noindent We have constructed a framework for economic theory based on the concept of generalized information equilibrium of \cite{Fielitz:2014} and used it to recover several macroeconomic toy models and show they are empirically accurate over post-war US economic data. A question that comes to the forefront: does the model work for other countries? The answer is generally yes\footnote{Some care is needed when looking at interest rates for e.g. the UK and Australia where foreign-currency denominated debt (in this case, US dollar) appears to cause countries to ``import'' the foreign interest rate.} (albeit with different model parameters), although a complete survey is ongoing (\cite{Smith:2015a}). Several examples appear in Figure \ref{fig:othersubfigures}.

This framework gives us a new perspective from which to interpret macroeconomic observations and tells us that sometimes macroeconomic effects are emergent and may not have microeconomic rationales\footnote{This does not mean they cannot be constructed as microeconomic interactions; they just do not need to be.}. Microfoundations, like Calvo pricing, may be an unnecessary theoretical requirement. However the information equilibrium may also be seen as satisfying the famous Lucas critique by utilizing information theoretic constraints to analyze empirical regularities in macroeconomic systems.

In general, the information equilibrium approach is agnostic about what mediates macroeconomic activity at the agent level or precisely how it operates. This may be unsatisfying for much of the field. However a useful analogy may be seen in physics. When Boltzmann developed statistical mechanics, the atoms he was describing -- although he believed they existed -- had not been established scientifically. The present approach can be thought of as looking at the economy from a telescope on a distant planet and treating economic agents as invisible atoms.

Even if it does not lead any further than the models presented here, the information equilibrium framework may still have a pedagogical use in standardizing and simplifying the approach to Marshallian crossing diagrams, partial equilibrium models and common classroom examples. A future paper \cite{Smith:2015b} will look into the connection between the utility maximization approach and an entropy maximization approach including: re-framing utility maximization as entropy maximization and interpreting the Euler equation and the asset pricing equation as maximum entropy conditions.

\section*{Acknowledgment}
\noindent We would like to thank Peter Fielitz, Guenter Borchardt and Tom Brown for helpful discussions and review of this manuscript.

\appendix

\section{Appendix}\label{app:modeldetails}

\noindent We have shown that several macroeconomic relationships and toy models can be easily represented using the information equilibrium framework, and in fact are remarkably accurate empirically. Below we list a summary of the information equilibrium models in the notation
\[
detector : source \rightleftarrows destination, 
\]
i.e. $ price : demand \rightleftarrows supply $. Also the information equilibrium models that do not require detectors are shown as 
\[
source \rightleftarrows destination.
\]
All data for the US is available at \cite{fred}, including the Solow model data for Mexico (real capital is inflated using the CPI less food and energy). The UK data is from the Bank of England website and FRED. The Japan data is from the Bank of Japan website and FRED. The Eurozone data is from the European Central Bank website and FRED. The models shown in Section \ref{s:macro} are: 
\vspace{0.5cm}

\textbf{AD-AS model}
\[
P: N \rightleftarrows S
\]

\textbf{Labor market (Okun's law)}
\[
P: N \rightleftarrows H \;\;\;\;\mbox{or}
\]
\[
P: N \rightleftarrows L
\]

Model parameters for the US
\begin{eqnarray}
k_{H} & = & 0.43 \mbox{ h/G\$} \nonumber
\end{eqnarray}

\textbf{IS-LM model}
\[
(i \rightleftarrows p ) : N \rightleftarrows M
\]
\[
i : N \rightleftarrows S
\]

Model parameters for the US interest rates (simultaneous fit)
\begin{eqnarray}
k_{i} & = & 3.49 \nonumber\\
k_{p} & = & 0.124 \nonumber
\end{eqnarray}

Model parameters for the UK interest rates (separate long, short fit)
\begin{eqnarray}
k_{i} & = & 2.71 \nonumber\\
k_{p} & = & 0.0344 \mbox{ (long)}\nonumber\\
k_{i} & = & 1.93 \nonumber\\
k_{p} & = & 0.0357 \mbox{ (short)}\nonumber
\end{eqnarray}

\textbf{Solow growth model}
\[
N \rightleftarrows K \rightleftarrows I
\]
\[
K \rightleftarrows D
\]
\[
N \rightleftarrows L
\]
\[
1/s : N \rightleftarrows I
\]
\[
(i \rightleftarrows p) : I \rightleftarrows M
\]

Model parameters for Mexico
\begin{eqnarray}
k_{1} & = & 0.51 \nonumber\\
k_{2} & = & 0.90 \nonumber\\
A & = & 0.0045 \nonumber
\end{eqnarray}

Model parameters for the US
\begin{eqnarray}
k_{1} & = & 0.44 \nonumber\\
k_{2} & = & 0.84 \nonumber\\
A & = & 0.0024 \nonumber
\end{eqnarray}

\textbf{Price level and inflation/quantity theory of money}
\[
P: N \rightleftarrows M
\]

Model parameters for the US, using the PCE price level $PCE(2009) = 1$
\begin{eqnarray}
M_{0} & = & 603.8 \mbox{ G\$}\nonumber\\
\alpha & = & 0.641 \nonumber\\
\gamma & = & 5.93 \times 10^{-4}\nonumber
\end{eqnarray}

Model parameters for Japan, using the core CPI price level 2010 index
\begin{eqnarray}
M_{0} & = & 12117.2 \mbox{ G\yen}\nonumber\\
\alpha & = & 0.673 \nonumber\\
\gamma & = & 1.17 \times 10^{-5}\nonumber
\end{eqnarray}

The definitions for the variables for all of these models are:
\begin{eqnarray}
N & & \mbox{nominal aggregate demand/output (NGDP)}\nonumber\\
M & & \mbox{monetary base minus reserves}\nonumber\\
H & & \mbox{total hours worked}\nonumber\\
L & & \mbox{total employed persons}\nonumber\\
S & & \mbox{aggregate supply}\nonumber\\
P & & \mbox{price level (core CPI or core PCE)}\nonumber\\
i & & \mbox{nominal long term interest rate (10-year rate)}\nonumber\\
p & & \mbox{price of money}\nonumber\\
K & & \mbox{nominal capital stock}\nonumber\\
D & & \mbox{nominal depreciation}\nonumber\\
I & & \mbox{nominal investment}\nonumber\\
s & & \mbox{savings rate}\nonumber
\end{eqnarray}

\section{Appendix}\label{app:codes}

\begin{figure}[h]
\begin{centering}
\includegraphics[width=0.7\columnwidth]{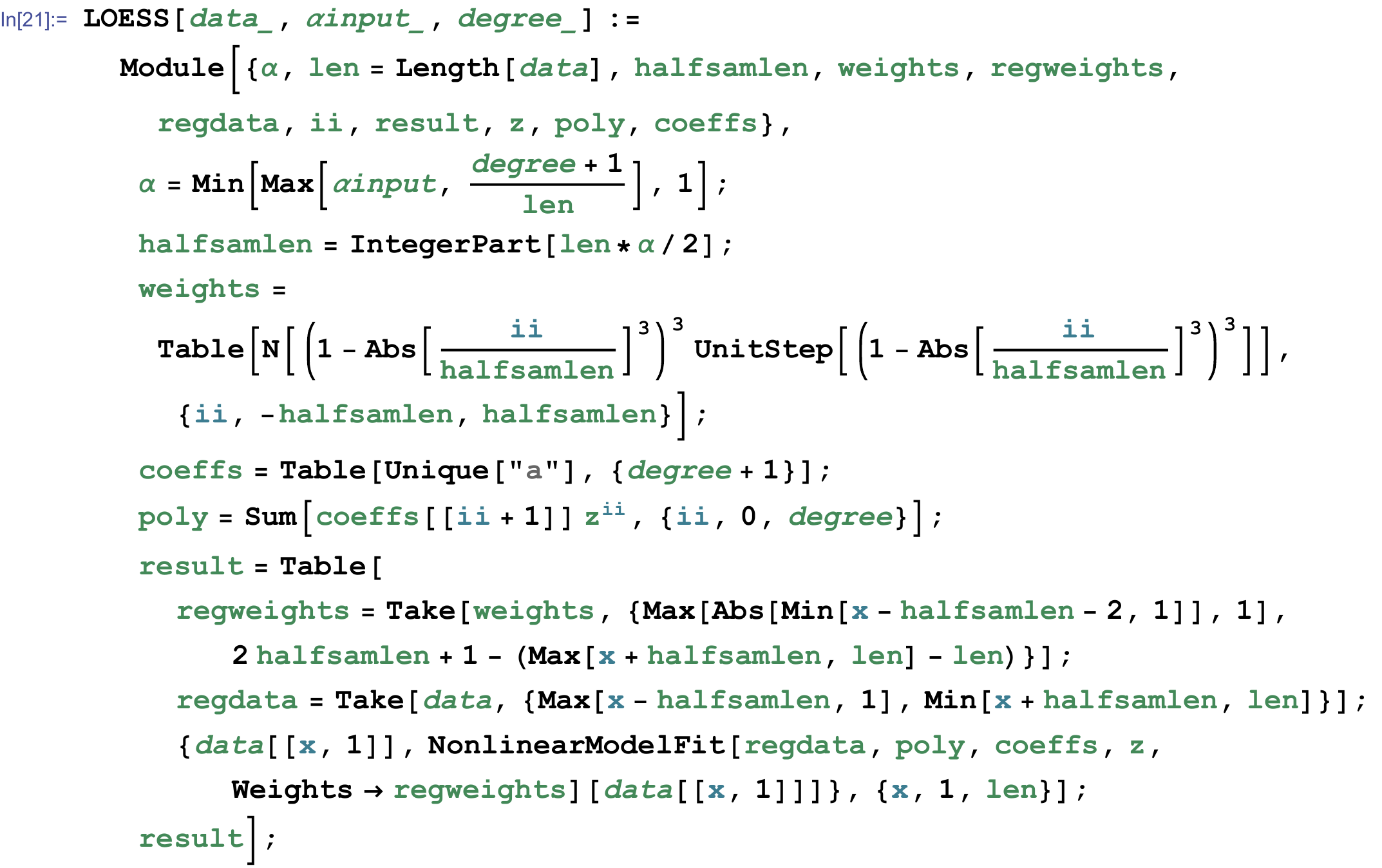}
\par\end{centering}
\caption{\label{fig:loess}\emph{Mathematica} code for performing LOESS smoothing.}
\end{figure}

In this appendix we show the numerical codes for the optimizations in Sections \ref{s:macro} and \ref{s:statecon}. They are written in \emph{Mathematica} using versions 8, 9 and 10. \emph{Mathematica} does not have its own local weighted regression (LOESS or LOWESS) smoothing function so we wrote one; the code is shown in Figure \ref{fig:loess}.

\noindent The parameter fits were accomplished by minimizing the residuals using the \emph{Mathematica} function $FindMinimum$ using the method $PrincipalAxis$, a derivative-free minimization method. The functions of the form $M0[yy]$ are a \emph{Mathematica} interpolating function with interpolation order set to linear using \cite{fred} data as input. $M0[yy]$ is the monetary base minus reserves (currency component), FRED series MBCURRCIR. $GDP[yy]$ is nominal gross domestic product FRED series GDP. $PCE[yy]$ is the personal consumption expenditures price level, excluding food and energy. $MB[yy]$ is the monetary base, FRED series AMBSL. $R03[yy]$ is the three month treasury bill secondary market interest rate, FRED series TB3MS.
\begin{figure}[h]
\begin{centering}
\includegraphics[width=0.7\columnwidth]{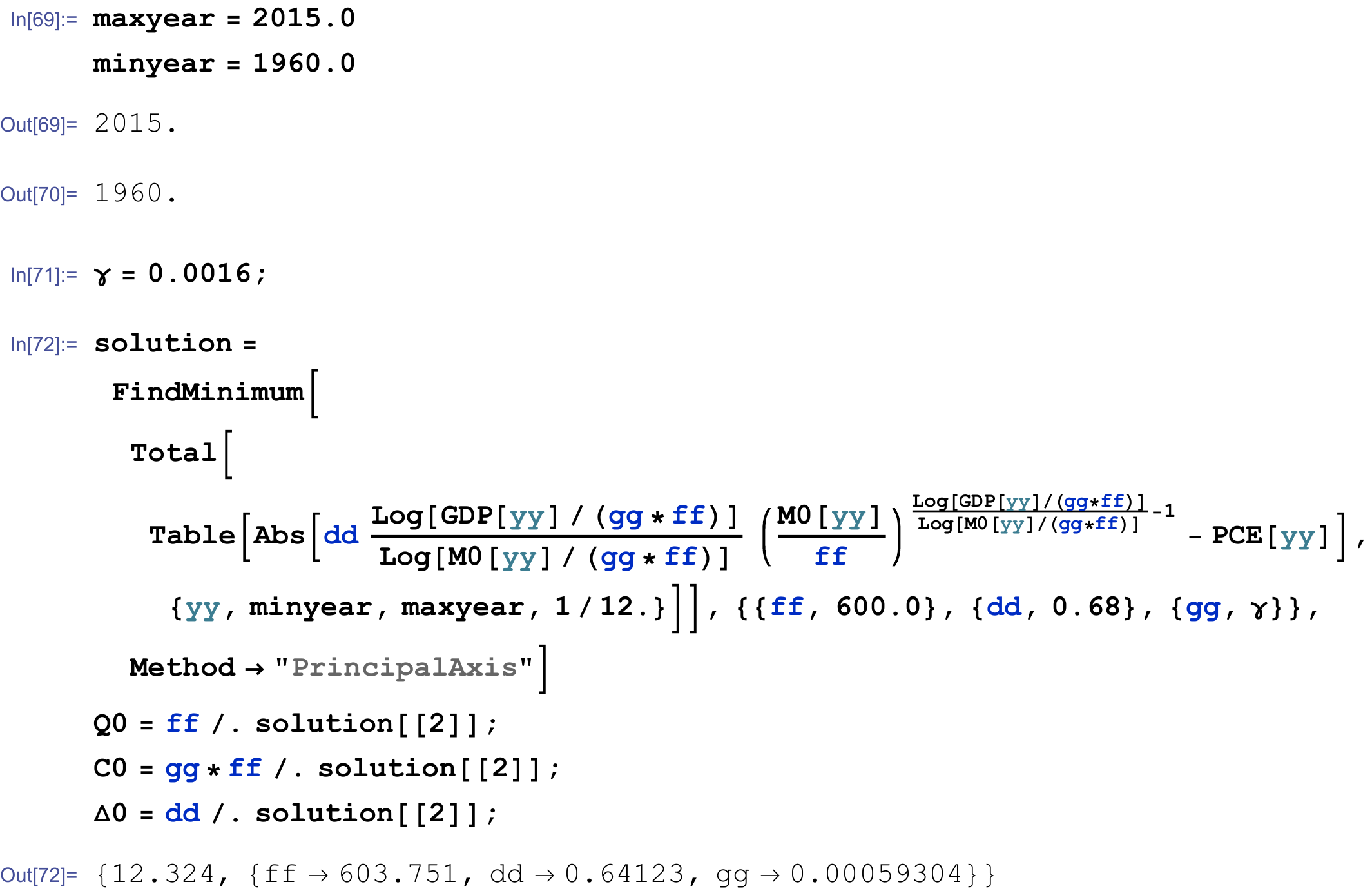}
\par\end{centering}
\caption{\label{fig:fitcode1}\emph{Mathematica} code for fitting the price level.}
\end{figure}
\begin{figure}[h]
\begin{centering}
\includegraphics[width=0.7\columnwidth]{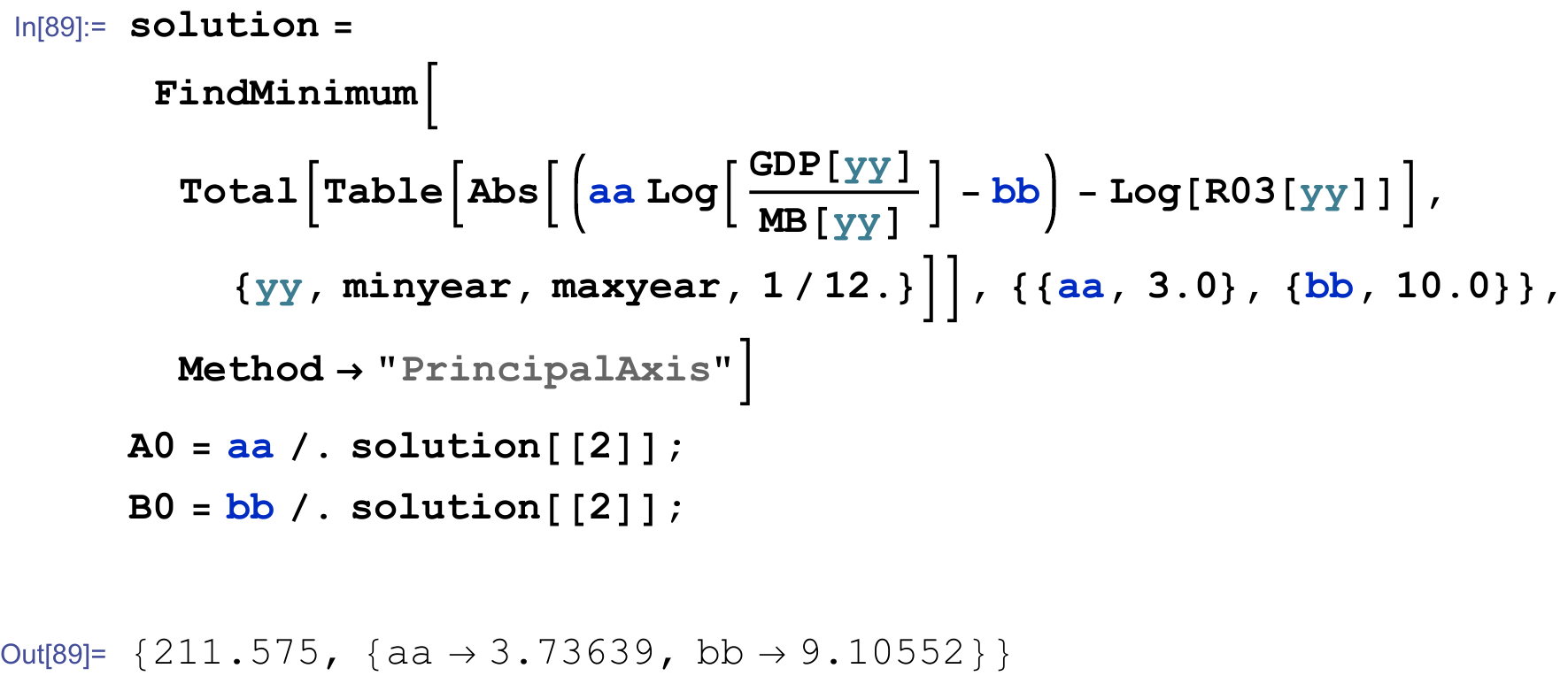}
\par\end{centering}
\caption{\label{fig:fitcode2}\emph{Mathematica} code for fitting the interest rate. The labor market model was fit using the similar code leaving out the parameter variable $aa$.}
\end{figure}

\begin{figure}[h]
\begin{centering}
\includegraphics[width=0.7\columnwidth]{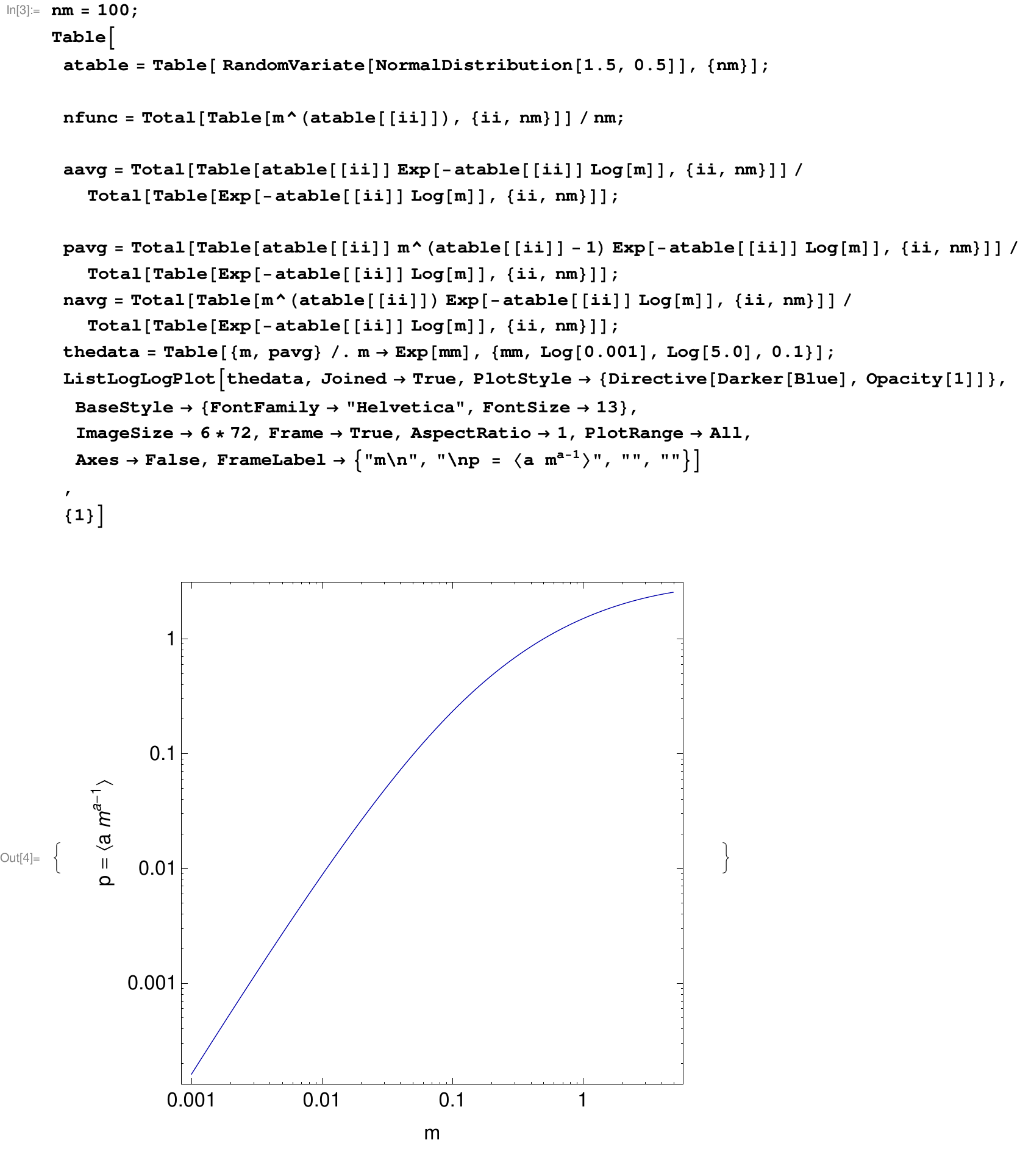}
\par\end{centering}
\caption{\label{fig:partitioncode1}Partition function calculation of the ensemble average of the price level. The 500 curves were generated by replacing $\{1\}$ with $\{500\}$.The different ensemble averages plotted the variables $aavg$, $navg$ and $pavg$.}
\end{figure}

\begin{figure}[h]
\begin{centering}
\includegraphics[width=0.7\columnwidth]{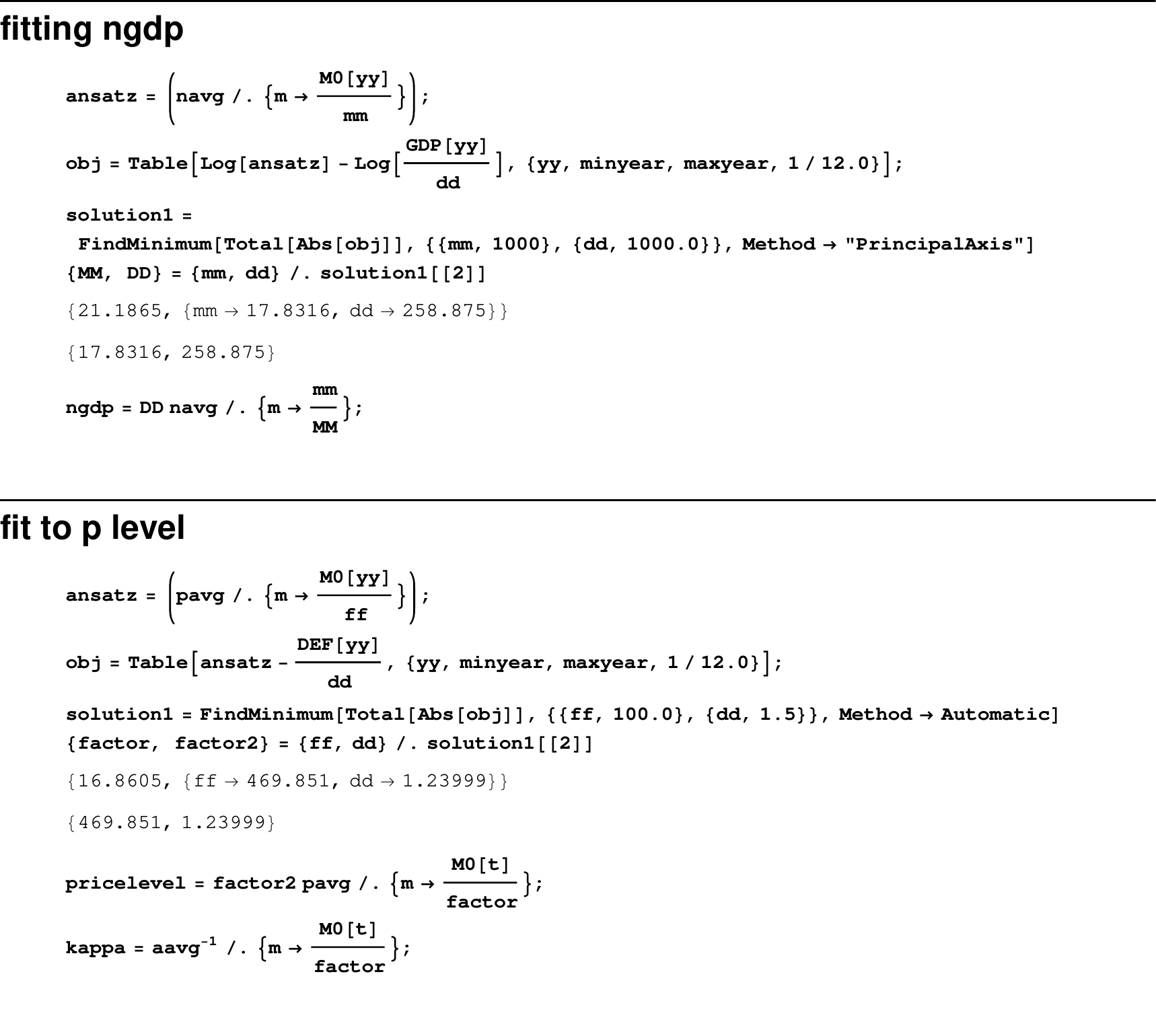}
\par\end{centering}
\caption{\label{fig:partitioncode2}Fitting the ensemble average to the price level and nominal output data.}
\end{figure}

\noindent Figures \ref{fig:partition1}, \ref{fig:partition2} and \ref{fig:partition3} in Section \ref{s:statecon} were generated with the code in Figure \ref{fig:partitioncode1}. The fits to the price level and nominal output used the code in Figure \ref{fig:partitioncode2}.

\section{Appendix}\label{app:sigmakappa}

\noindent If we keep the parameter $\gamma$ constant across countries, it can aid cross-national comparisons as we show in this appendix. First, set up the variables 
\begin{eqnarray}
\kappa & = & 1/k(N,M) = \frac{\log M/C_{0}}{\log N/C_{0}}\\
\sigma & = & \frac{M}{M_{0}}
\end{eqnarray}
setting up the constant $C_{0}$. I call these the information transfer index (from the original theory) and the normalized monetary base, respectively. Defining the constant
\[
\alpha = \frac{N_{0}}{M_{0}}
\]
we can write
\[
P = \alpha \frac{1}{\kappa} \sigma^{1/\kappa-1}
\]

Calculating the derivative above (after dividing by $\alpha$), one obtains 
\[
\frac{\partial P(\kappa,\sigma)}{\partial\sigma} =\frac{\partial}{\partial\sigma}\frac{\log N/C_0}{\log \sigma M_{0}/C_0} \sigma^{\frac{\log N/C_0}{\log \sigma M_{0}/C_0}-1} = 0 \]
\[\frac{P(\kappa, \sigma)}{\sigma} \left[ \frac{\log N/C_0}{\log \sigma M_{0}/C_0} \left(\frac{\log \sigma }{\log \sigma M_{0}/C_0}-1\right)+\log \sigma M_{0}/C_0 + 1 \right] = 0
\]
The bracketed term must be zero since the piece outside the bracket is positive, so therefore, after some substitutions 
\[
-\frac{1}{\kappa}\log \left(\frac{M_{0}}{C_0}\right)+\log \left(\frac{\sigma M_{0}}{C_0}\right)+1=0
\]
And we arrive at
\[
\sigma = \frac{C_0}{ M_{0}} \exp \left( -\frac{\kappa + \log \frac{C_0}{ M_{0}}}{\kappa} \right)
\]
Note that this function is only of $\sigma$, $\kappa$ and $C_0/M_{0} \equiv \gamma$. This means if we use the parameters for one country to find $\gamma$, we can then constrain the subsequent fits for Japan (and other countries) to maintain $C_0 = \gamma M_{0}$ (reducing one degree of freedom). This constrains the fits so that the ridge lines where $\partial P/\partial M = 0$ coincide.

\section{Appendix}\label{app:equiv}

\noindent Information equilibrium is an equivalence relation. If we define the statement $A$ to be in information equilibrium with $B$ (which we'll denote $A \rightleftarrows B$) by the relationship (i.e. ideal information transfer between $A$ and $B$):
\begin{equation}
\frac{dA}{dB} = k \frac{A}{B}
\end{equation}
for some value of $k$, then, first we can show that $A \rightleftarrows A$ because
\begin{eqnarray}
\frac{dA}{dA} & = & k \frac{A}{A}\\
1 & = & k \cdot 1
\end{eqnarray}
and we can take $k = 1$. Second we can show that $A \rightleftarrows B$ implies $B \rightleftarrows A$ by re-deriving the relationship \ref{eq:main}, except moving the variables to the opposite side:
\begin{equation}
\frac{dB}{dA} = \frac{1}{k}\;\; \frac{B}{A} = k' \; \frac{B}{A}
\end{equation}
for some $k'$ (i.e. $k' = 1/k$). Lastly we can show that $A \rightleftarrows B$ and $B \rightleftarrows C$ implies $A \rightleftarrows C$ via the chain rule:
\begin{eqnarray}
\frac{dA}{dB} & = & a \frac{A}{B}\\
\frac{dB}{dC} & = & b \frac{B}{C}
\end{eqnarray}
such that
\begin{equation}
\frac{dA}{dC} = \frac{dA}{dB}\; \frac{dB}{dC} = a b \; \frac{A}{B} \frac{B}{C}
\end{equation}
\begin{equation}
\frac{dA}{dC} = k \; \frac{A}{C}
\end{equation}
with information transfer index $k = a b$. That gives us the three properties of an equivalence relation: reflexivity, symmetry and transitivity.

\end{document}